%% file: main_rev_2.tex
\documentclass[10pt,journal,compsoc]{IEEEtran}

\ifCLASSOPTIONcompsoc
  \usepackage[nocompress]{cite}
\else
  \usepackage{cite}
\fi

\usepackage[english]{babel}
\usepackage{blindtext}
\usepackage{tikz,pgfplots}
\usepackage{tkz-euclide}
\pgfplotsset{compat=1.5}
\usetikzlibrary{calc}
\usepgfplotslibrary{polar}

\usepackage{epstopdf} 
\usepackage{amsmath}
\usepackage{amsfonts}
\usepackage{wrapfig}
\usepackage{mathrsfs}
\usepackage{accents}
\usepackage{acronym}
\usepackage{graphicx}
\usepackage{textcomp}
\usepackage{xcolor}
\usepackage{mathtools}
\usepackage{color,soul, colortbl}
\usepackage{graphicx}
\usepackage{booktabs}
\usepackage{multirow}
\usepackage[font={footnotesize}]{caption}
\usepackage[font={footnotesize}]{subcaption}
\usepackage{algorithm}
\usepackage{algorithmic}
\usepackage[shortlabels]{enumitem}

\DeclareMathOperator*{\argmax}{arg\,max}
\DeclareMathOperator*{\argmin}{arg\,min}

\newcommand{\smallsection}[1]{\noindent\textbf{#1.}}

\usepackage{caption}
\usepackage{subcaption}
\usepackage{url}

\newcommand{\eq}[1]{Eq.~\eqref{#1}}
\newcommand{\fig}[1]{Fig.~\ref{#1}}
\newcommand{\tab}[1]{Tab.~\ref{#1}}
\newcommand{\secref}[1]{Section~\ref{#1}}

\input{ACRONYMS.tex}

\newcommand{\rev}{\textcolor{blue}} 
\renewcommand{\rev}{}

\usepackage{scalerel}
\usepackage{tikz}
\usetikzlibrary{svg.path}

\definecolor{orcidlogocol}{HTML}{A6CE39}
\tikzset{
  orcidlogo/.pic={
    \fill[orcidlogocol] svg{M256,128c0,70.7-57.3,128-128,128C57.3,256,0,198.7,0,128C0,57.3,57.3,0,128,0C198.7,0,256,57.3,256,128z};
    \fill[white] svg{M86.3,186.2H70.9V79.1h15.4v48.4V186.2z}
                 svg{M108.9,79.1h41.6c39.6,0,57,28.3,57,53.6c0,27.5-21.5,53.6-56.8,53.6h-41.8V79.1z M124.3,172.4h24.5c34.9,0,42.9-26.5,42.9-39.7c0-21.5-13.7-39.7-43.7-39.7h-23.7V172.4z}
                 svg{M88.7,56.8c0,5.5-4.5,10.1-10.1,10.1c-5.6,0-10.1-4.6-10.1-10.1c0-5.6,4.5-10.1,10.1-10.1C84.2,46.7,88.7,51.3,88.7,56.8z};
  }
}

\newcommand\orcidicon[1]{\href{https://orcid.org/#1}{\mbox{\scalerel*{
\begin{tikzpicture}[yscale=-1,transform shape]
\pic{orcidlogo};
\end{tikzpicture}
}{|}}}}

\usepackage[hidelinks]{hyperref} 
\usepackage{cleveref}

\begin{document}

\title{RAPID: Retrofitting IEEE 802.11ay Access Points for Indoor Human Detection and Sensing
\thanks{$^{\S}$ Corresponding author e-mail: \texttt{pegoraroja@dei.unipd.it}
\newline $^{\dag}$ These authors are with the Department of Information Engineering, University of Padova, Italy.\newline
$^*$ These authors are with the IMDEA Networks Institute, 28918 Madrid, Spain. \newline
This research work was supported by the European Union’s Horizon 2020 research and innovation programme under grants No. 871249: ``LOCalization and analytics on-demand embedded in the 5G ecosystem for Ubiquitous vertical applicationS'' (LOCUS) and No. 861222: ``MIllimeter-wave NeTworking and Sensing for beyond 5G'' (MINTS), by the Spanish Ministry of Science and Innovation (MICIU) grant RTI2018-094313-B-I00 (PinPoint5G+), by the Region of Madrid through TAPIR-CM (S2018/TCS-4496) and by the Italian Ministry of Education, University and Research (MIUR) through the initiative “Departments of Excellence” (Law 232/2016).
}
}
\author{
Jacopo Pegoraro$^{\dag \S}$\textsuperscript{\orcidicon{0000-0003-3555-5666}}, 
Jesus O. Lacruz$^*$\textsuperscript{\orcidicon{0000-0002-6641-2003}}, 
Francesca Meneghello$^\dag$\textsuperscript{\orcidicon{0000-0002-9905-0360}}, \\
Enver Bashirov$^\dag$\textsuperscript{\orcidicon{0000-0002-3524-1088}}, 
Michele Rossi$^\dag$\textsuperscript{\orcidicon{0000-0003-1121-324X}},
and 
Joerg Widmer$^*$\textsuperscript{\orcidicon{0000-0001-6667-8779}} 
\vspace{-0.5cm}}

\markboth{IEEE Transactions on Mobile Computing,~Vol.~XX, No.~XX, Month~Year}%
{Pegoraro \MakeLowercase{\textit{et al.}}: RAPID: Retrofitting IEEE 802.11ay Access Points for Indoor Human Detection and Sensing}
\markboth{}%
{Pegoraro \MakeLowercase{\textit{et al.}}: RAPID: Retrofitting IEEE 802.11ay Access Points for Indoor Human Detection and Sensing}

\IEEEtitleabstractindextext{%
\begin{abstract}
In this work we present RAPID, the first joint communication and radar system based on next-generation IEEE~802.11ay WiFi networks operating in the $60$~GHz band. Unlike existing approaches for human sensing at millimeter-wave frequencies, which rely on special-purpose radars, RAPID achieves radar-level sensing accuracy with IEEE~802.11ay access points, thus avoiding the burden of installing ad-hoc sensors. 
RAPID enables contactless human sensing applications, such as people tracking, Human Activity Recognition (HAR), and person identification 
without requiring modifications to the standard packet structure. 
Specifically, we leverage IEEE~802.11ay beam training to accurately localize and track multiple individuals within the same environment. Then, we propose a new way of using beam tracking to extract micro-Doppler signatures from the time-varying Channel Impulse Response (CIR) estimated from \textit{reflected} packets. Such signatures are fed to a deep learning classifier to perform HAR and person identification.
RAPID is implemented on a cutting-edge IEEE~802.11ay-compatible FPGA platform with phased antenna arrays, and evaluated on a large dataset of CIR measurements. It is robust across different environments and subjects, and outperforms state-of-the-art sub-$6$~GHz WiFi sensing techniques. Using two access points, RAPID reliably tracks multiple subjects, reaching HAR and person identification accuracies of $94\%$ and $90\%$, respectively.
\end{abstract}

}

\maketitle

\IEEEdisplaynontitleabstractindextext

\IEEEpeerreviewmaketitle

\section{Introduction}
\label{sec:intro}

In this work, we design RAPID, a pervasive \ac{jcr} system that extends the capabilities of upcoming WiFi technology operating in the $60$~GHz \ac{mmwave} spectrum to integrate sensing functionalities into wireless networks. The joint provisioning of communication and sensing services is of great value to pave the way toward advanced smart-home applications without the need for deploying dedicated sensing hardware. In this regard, our target is to retrofit IEEE~802.11ay hardware so as to natively offer human and environment sensing services to end users, in addition to high-throughput communication.

Thanks to their large available bandwidth, \ac{mmwave} signals allow performing localization and tracking with decimeter-level accuracy, making them the preferred solution for contactless sensing through radio waves. Most emerging \ac{mmwave} sensing systems are based on \textit{dedicated} \ac{mmwave} radar devices and estimate the \ac{md} effect induced by human motion (\textit{signature}) with high accuracy via specifically designed bursts of phase coherent chirp signals \cite{vandersmissen2018indoor, seifert2019toward}. Radar \ac{md} signatures contain detailed information about the movement velocity of the different human body parts across time, and enable fine-grained sensing applications such as person identification from gait features \cite{vandersmissen2018indoor, pegoraro2021multiperson, meng2020gait, zhao2019mid}, \ac{har} \cite{singh2019radhar}, gait disorder diagnosis \cite{seifert2019toward} and fall detection \cite{jin2020mmfall}, among others. However, solutions based on \ac{mmwave} radars come with the drawback of the need for installing bespoke sensors, which limits their scalability and ease of deployment in practical scenarios (e.g., smart buildings, offices, etc.).

In this respect, the ubiquitous deployment of WiFi devices has sparked research interest towards developing joint communication and \ac{rf} sensing technology, to avoid the cost of installing dedicated hardware while at the same time benefiting from communication capabilities. The effort of enhancing WiFi devices with environment sensing features has recently led to the establishment of the IEEE~802.11bf standardization group \cite{chen2022wi}, aimed at integrating sensing functionalities into WiFi-enabled devices. While legacy WiFi technology based on IEEE~802.11n/ac/ax (sub-$6$~GHz bands) standards provides a viable means for environment and human sensing \cite{ma2019wifi} and \ac{har} \cite{wang2017device,chen2018wifi}, it suffers from intrinsic limitations due to the relatively low bandwidth available in the sub-$6$~GHz license-exempt portion of the radio spectrum. This prevents highly accurate distance measurements and multi-person localization and tracking in realistic scenarios. Moving to the \ac{mmwave} spectrum, previous works based on the IEEE~802.11ad standard exploit the \ac{cir} estimation procedure for localizing people \cite{wu2020mmtrack, zhang2020mmeye}, but they are not fully compliant with the communication packet structure specified by the standard and cannot match the sensing accuracy of radars, as no \ac{md} information is captured. 
Overall, the extraction of \ac{md} signatures is difficult using standard communication devices and protocols, due to the lack of specifically designed waveforms and transmission modes. Extracting Doppler information from sequences of subsequent packets, as done in radars, is highly non-trivial due to the random and time varying phase offsets between the transmitter and the receiver \cite{zhang2021overview}. In fact, the offsets destroy the phase coherence across different packets, preventing the extraction of \ac{md} signatures which require a phase analysis across long sequences of subsequently transmitted signals.

RAPID is the first system that successfully extracts \ac{md} signatures of human movements using standard WiFi transmission technology working on the \ac{mmwave} spectrum, and achieves radar-level accuracy in sensing. 
It works \textit{without modifying the packet structure} by leveraging the \textit{in-packet} beam training and beam tracking features of IEEE~802.11ay. This leads to very low implementation and deployment cost, and allows for a highly accurate extraction of human movement information from the radio signals.
 
IEEE~802.11ay uses highly directional antennas to shape precise beams for communication. For that, the standard specifies efficient in-packet beam training and tracking procedures \cite{Knightly_ICOMM2017}, based on training (TRN) fields consisting of repetitions of complementary Golay sequences \cite{Lacruz_MOBISYS2021}. The fields are transmitted with different beam patterns, which allows determining which of them is best for communication. 
By exploiting \textit{beam training} packets, RAPID accurately localizes multiple human subjects within the same indoor space. Then, the \ac{md} signature associated with the movement of each subject is extracted by relying on the TRN units embedded in the data packets used for \textit{beam tracking}, analyzing the phase differences of the \ac{cir} across subsequent packets that are reflected back by the environment. The obtained \ac{md} spectrograms are processed using deep learning classifiers to carry out continuous \ac{har} and person identification.

Thanks to the intrinsic superior ranging resolution of the \ac{mmwave} spectrum and our advanced signal processing, RAPID outperforms state-of-the-art human sensing technology based on sub-$6$~GHz WiFi systems. RAPID allows individually tracking multiple moving subjects, separating their signal reflections and, in turn, obtaining large improvements in terms of accuracy, robustness and generalization across environments and subjects. In addition, multiple RAPID-\acp{ap} can be seamlessly integrated to boost detection and tracking performance. This also increases \ac{har} and person identification accuracy by combining the information from different viewpoints.

In this work, RAPID-\acp{ap} are implemented using an FPGA-based \ac{sdr} platform equipped with phased antenna arrays, which transmits IEEE~802.11ay-compliant packets and operates in a full-duplex fashion.
RAPID IEEE~802.11ay \acp{ap} enable their transmit and receive chains simultaneously, avoiding the problem of random phase offsets as transmitter and receiver share the same local oscillator. Note that this does not require complex self-interference cancellation for full-duplex communication, since the receiver needs to only detect the highly robust Golay sequences of the TRN fields.

We stress that RAPID is not simply about applying radar signal processing to a different domain. Reusing standard-compliant IEEE~802.11ay signals requires developing new processing steps to obtain range, angle, and \ac{md} information, while taking care of \ac{jcr}-specific problems that do not arise in radar systems. 
While radars typically estimate the channel using ad-hoc \textit{chirp} waveforms, whose parameters can be \textit{tuned} to meet the specific sensing requirements, RAPID re-uses standard-compliant Golay sequences. Therefore, the sensing resolution can not be adapted to the considered scenario, and the person detection and range estimation steps have to be entirely re-designed to be robust under such constraints.  
For what concerns the \ac{aoa} estimation, \ac{mmwave} radars are usually equipped with \ac{mimo} antenna arrays that ease the estimation of the \ac{aoa} by analyzing the phase change across the spatial dimension. On the contrary, IEEE~802.11ay \acp{ap} typically mount cheaper phased array antennas, therefore a different approach has to be designed to obtain the \ac{aoa} by analyzing the \ac{cir} estimated through different beam patterns (\secref{sec:aoa-est}). 
Lastly, the \ac{md} computation is challenging as it involves \textit{(i)} striking a good balance between the packet transmission rate and the Doppler frequency resolution required to capture the \ac{md} of human movement, while \textit{(ii)} ensuring sufficient phase coherence across adjacent packets (see \secref{sec:md-sep}). Moreover, Golay sequences are known to have low Doppler resolution \cite{Kumari_11adradar_VTC}, and no existing study has evaluated the feasibility of using them to extract fine-grained human \ac{md} signatures.

\setlist[enumerate,1]{leftmargin=0.4cm}
To summarize, the main contributions of our work are:\vspace{-0.08cm}
\begin{enumerate}
    \item We design and implement RAPID, a fully standard compliant \ac{jcr} system that exploits IEEE~802.11ay TRN fields to achieve radar-like human sensing, including simultaneous multi-person tracking, \ac{har} and person identification. RAPID reuses existing fields in the communication packets and avoids the need for a dedicated sensing infrastructure. RAPID can leverage data from a single \ac{ap} or combine information from multiple \acp{ap} for improved performance. 
    \item We propose a novel method to extract \ac{md} signatures of human movement from IEEE~802.11ay \ac{cir} estimates obtained from a sequence of IEEE~802.11ay data packets with added beam tracking fields, exploiting the Golay sequences specified in the standard. To the best of our knowledge, this is the first work to do so.
    \item We implement RAPID on a novel FPGA-based testbed including multiple IEEE~802.11ay-compliant \acp{ap} which support full-duplex operation, so that each \ac{ap} can listen to its own transmitted signal and act as a monostatic \ac{jcr} device.
    \item We conduct an extensive indoor measurement campaign to evaluate the proposed system and compare it to sub-$6$~GHz WiFi systems. To this end, we build a unique dataset including simultaneous IEEE~802.11ay and IEEE~802.11ac \ac{cir} estimates. RAPID achieves continuous tracking of up to $5$ concurrently moving subjects, with \ac{har} accuracy of $94\%$ and person identification accuracy of $90\%$. Moreover, it outperforms state of the art sub-$6$~GHz WiFi sensing, showing superior accuracy and robustness to different environments and subjects. 
\end{enumerate}

The paper is organized as follows. The related work is summarized in \secref{sec:rel}. RAPID is introduced in \secref{sec:rapid}, presenting its constituent processing blocks. A summary of how IEEE~802.11ay can be used for environment sensing is given in \secref{sec:11ay}, while in \secref{sec:imp} the implementation of RAPID on FPGA hardware is discussed. A thorough performance analysis of RAPID on real measurements is presented in \secref{sec:ex-res}. \secref{sec:conclusion} concludes the discussion.

\section{Related work}\label{sec:rel}

\smallsection{Sub-6~GHz sensing} Legacy WiFi technologies such as IEEE 802.11n and IEEE 802.11ac, respectively working at $2.4$ or $5$~GHz, have been extensively used for human sensing, including activity/gesture recognition~\cite{li2019wi, chen2018wifi, wang2017device, meneghello2021environment}, vital sign monitoring~\cite{wang2017phasebeat} and person identification~\cite{korany2020multiple}. 
Due to the rich multipath environment at lower frequencies, existing approaches have reached good accuracies by leveraging OFDM transmission and analyzing the \ac{cir} amplitude obtained at the different subcarriers, as done in~\cite{wang2017device}. The performance of such systems can be further improved by exploiting the phase components of the \ac{cir}~\cite{meneghello2021environment,li2019wi}, but this entails using complex algorithms for the removal of random phase offsets.

Although there is a large body of work that exploits these technologies, they have two main drawbacks: \textit{(i)} they are effective for single-person scenarios, as the small available bandwidth only allows for coarse localization and tracking of the subjects, and \textit{(ii)} they are highly sensitive to changes in the environment and hardly generalize to new scenarios (never seen at system calibration/training time), which can significantly worsen their performance. Addressing problem \textit{(i)}, in \cite{korany2020multiple}, multi-person identification using IEEE~802.11n is achieved in a through-the-wall setting, but the subjects still need to be well separated in space (e.g., by at least $20^{\circ}$ in azimuth angle at a distance of several meters). To mitigate the dependence on the environment, more elaborate deep learning and optimization approaches have been proposed in~\cite{jiang2018towards, shi2022environment, meneghello2021environment}. However, they still are not able to approach radar-like sensing accuracy.
\smallskip

\smallsection{\ac{mmwave} radar sensing} \ac{mmwave} frequencies offer a natural solution to the above issues, by providing decimeter-level accuracy in distance measurements and high sensitivity to the \ac{md} effect, due to their small transmission wavelength. In addition, due to the sparsity of the \ac{mmwave} channel, higher robustness to environmental changes is achieved. 
\ac{mmwave} radars have been intensively studied in the past few years as an effective means to achieve fine grained environment sensing \cite{davoli2021machine}. Typical operating frequencies for these devices are the $60$ or the $77$~GHz bands. 
Centimeter-level accuracy in measuring distances is achieved thanks to the use of very large transmission bandwidths, up to $4$~GHz, as dedicated radar devices are not constrained by communication requirements.
Radars allow accurate \ac{har} \cite{singh2019radhar, lai2021radar} and have been used to perform person identification on small to medium-sized groups of people (up to a few tens), due to their very high resolution in obtaining the \ac{md} signatures of the subjects \cite{meng2020gait, pegoraro2021realtime}. In these works, the separation of the reflections from subjects concurrently moving in the environment is achieved through \ac{mimo} radars, which enable high angular resolution and allow tracking the users with errors below $0.2$~m even in realistic scenarios where people walk and move freely \cite{zhao2019mid}. However, these results are obtained within relatively small distances from the radar, ranging from $4$ \cite{zhao2019mid} to $6-7$~m \cite{meng2020gait}.

Despite the advanced sensing capabilities, \ac{mmwave} radars entail high deployment costs to cover large indoor areas, even more considering their limited working range. For this reason, multi-radar networks to cover wider areas and avoid occlusions are seldom considered in the literature. Reusing existing \ac{mmwave} communication links, as we do in this work, allows avoiding the costly deployment of additional hardware, while maintaining radar-like human sensing and detection performance.
\smallskip

\smallsection{802.11ad $\boldsymbol{60}$~GHz sensing} Commodity $60$~GHz radios have been utilized for client device localization \cite{pefkianakis2018accurate}, people tracking \cite{wu2020mmtrack}, fine-grained human gesture recognition \cite{regani2021mmwrite, ren2021hand}, vital sign monitoring \cite{wang2020vimo} and \ac{rf} imaging \cite{zhang2020mmeye}.
Among them, in \cite{regani2021mmwrite}, pulsed radar-like operations are performed to detect and track a human hand, reconstructing handwriting with centimeter-level accuracy. Notably, \cite{ren2021hand} performs similar processing using the IEEE~802.11ad \ac{cir} estimated by a mobile device for gesture classification.
In \cite{zhang2020mmeye}, a commodity $60$~GHz radio equipped with a $6\times 6$ antenna array is used to obtain the silhouette of a person moving directly in front of the device. This is achieved with an angular super-resolution algorithm derived from MUSIC \cite{schmidt1986multiple}. However, the device needs to be operated in a \textit{radar mode} for transmission, which may not comply with the communication standard.
In \cite{wu2020mmtrack}, the estimated \ac{cir} amplitude is used along with receiver beamforming to localize and track multiple people, achieving a median localization error of $9.9$~cm. This work does not exploit the phase of the \ac{cir} to extract the \ac{md} signature of the subjects, which is necessary to carry out \ac{har} and person identification tasks. Moreover, the extension to the case of multiple \ac{ap}s is not considered.
Overall, the research addressing human sensing through the IEEE~802.11ad standard typically does not consider the \textit{joint} communication and sensing aspect, which requires to reuse the packet structure specified by the communication standard.
\smallskip

\smallsection{802.11ay $\boldsymbol{60}$~GHz sensing} To the best of our knowledge, RAPID is the first system that extracts \mbox{radar-like} \ac{md} signatures of human movement from IEEE~802.11ay $60$~GHz \ac{ap}s, by retrofitting them with human sensing and \ac{md} extraction capabilities. This is obtained by preserving the IEEE~802.11ay packet structure, thus obtaining a \textit{joint radar-communication platform that is fully standard compliant}. 

\section{RAPID sensing system}
\label{sec:rapid}

\begin{figure*}[t!]
	\begin{center}   
		\includegraphics[width=15cm]{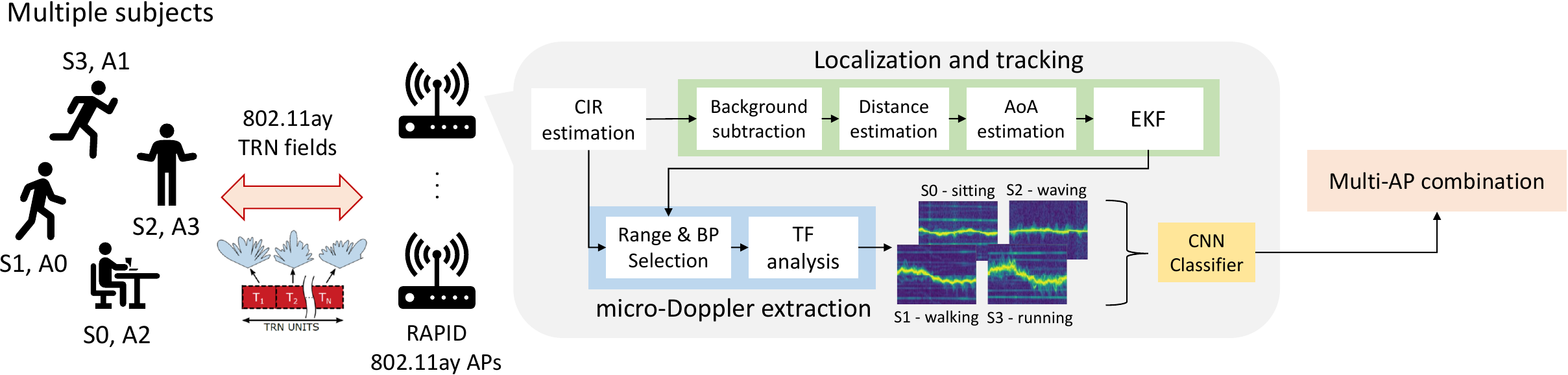} 
		\caption{Overview of the RAPID system.
		}
		\label{fig:workflow}
	\end{center}
 \vspace{-0.4cm}
\end{figure*}
RAPID enables indoor human sensing in IEEE~802.11ay networks, by leveraging the network \textit{in-packet} beam training and beam tracking fields. In the following, the system is presented by detailing the processing blocks that allow performing people localization and tracking, \ac{har}, and person identification. The novel algorithms specifically designed to extract range, \ac{aoa} and Doppler estimates from the \ac{cir} obtained though standard-compliant Golay sequences are deepened. The mathematical models of the \ac{cir} and the \ac{cir} phase are included to make the analysis self-comprehensive.

\subsection{System overview}
\label{sec:overview}

From a high-level perspective, RAPID performs the following operations, as shown in \fig{fig:workflow}.

    \noindent \textbf{(1) IEEE~802.11ay \ac{cir} estimation:} 802.11ay specifies the transmission of a variable number of TRN units for in-packet beam training, each using a (possibly) different \ac{bp}. From the \ac{cir}, which is estimated from each TRN unit (see \secref{sec:11ay}), RAPID obtains a scan of the whole angular \ac{fov}, which contains accurate information about all the surrounding objects and people.\\
    \noindent \textbf{(2) People localization and tracking:} the individuals are detected by performing background subtraction from the \ac{cir} amplitude and applying a thresholding algorithm to detect candidate reflection paths from humans, see \secref{sec:back-sub} and \secref{sec:dist-est}, respectively. Subsequently, a correlation based algorithm is utilized to estimate the angular position of the subjects, as described in \secref{sec:aoa-est}, and an \ac{ekf} is exploited to sequentially track and refine the positions of the individuals across time (\secref{sec:ekf}). By combining more than one \ac{ap}, RAPID can boost its human detection capabilities, while effectively coping with occlusion problems, as quantified in \secref{sec:prel-exp}.\\
    \noindent \textbf{(3) $\boldsymbol{\mu}$D spectrum extraction:} here, the \ac{md} spectrum of each detected person is extracted. This is implemented by utilizing the \ac{cir} model as a radar return signal, and using the estimated positions from point $(2)$ to single out the \ac{cir} portions (the paths and the \ac{bp}s) containing the contributions of each subject, see \secref{sec:md-sep}. The \ac{md} signature of each individual's movement is then extracted by computing the power spectrum of the corresponding \mbox{complex-valued} portion of the \ac{cir} over windows of suitable length, employing \ac{tf} analysis.\\
    \noindent \textbf{(4) \ac{har} and person identification:} the spectrograms from step $(3)$ are fed to a deep learning classifier based on a residual \ac{cnn} \cite{he2016deep} for \ac{har}. Thanks to the separation of the \ac{cir}, and to the subsequent computation of the \ac{md} for each individual, RAPID is capable of recognizing the different activities performed by multiple subjects within the same indoor space. Moreover, through a second \ac{cnn} module, it is also able to identify a person, by extracting and analyzing their gait features from the \ac{md} signature. With multiple \ac{ap}s, the classifications are refined by selecting the best \ac{ap} to make the decision, according to the confidence of the classifier output.


In this work, we aim at localizing and tracking people within a given physical space, by identifying which person is performing which activity. This requires person identification, tracking and \ac{har} capabilities. The person identification task is carried out by extracting and analyzing the \ac{md} associated with the human gait, as this is an effective (soft) biometric signature, which has been successfully used in many works \cite{nambiar2019gait}. Hence, we first detect when a person is walking, then we get his/her identity from the \ac{md} gait signature and, finally, we keep tracking the person by also recognizing their activities. This also works the other way around, i.e., if a person is at first sitting and doing other activities, and then starts walking later on; as long as tracking works, we can later determine who was sitting earlier on. This also explains why tracking a person is critical, so that it is still clear which person is where, even when he/she performs other activities than walking. 


We now present in detail each RAPID processing function, following the workflow of \fig{fig:workflow}.

\subsection{\ac{cir} estimation}
\label{sec:cir-model}

\ac{cir} estimation is a key component of most communication systems and is used to obtain information about the environmental reflections of the signal -- such as their associated angle of arrival and delay at the receiver -- to properly set the data transmission parameters and decode received packets. RAPID leverages this process for sensing purposes. A key aspect to our design is that the large transmission bandwidth of \ac{mmwave} systems leads to \ac{cir} containing fine-grained information about the environment. In our system, the transmitter and the receiver units are co-located: the signal sent by the former, after bouncing off nearby reflectors (objects or humans), is collected at the receiver that retrieves information for each reflector, such as its distance and angular position with respect to the device, its moving velocity and micro-Doppler. 

The \ac{cir} is represented as a vector of complex channel gains, also referred to as \textit{paths} in the following, and indicized through letter $\ell$. 
Due to the finite delay resolution of the system, the \ac{cir} vector can only represent a discrete grid of paths, with corresponding propagation delays $\tau_{\ell} = \ell / B,  \ell=0, \dots, L-1$, where $B$ is the transmission bandwidth and $1/B$ is the delay resolution.
The components of the \ac{cir} vector, which represent the complex gains for the $L$ paths, are obtained by correlating the received signal with pre-defined Golay sequences, using standard techniques \cite{Garcia_INFOCOM2020}, see also \secref{sec:11ay}.  
Path $\ell$ is mapped onto the corresponding reflector distance using $d_{\ell}=c\tau_{\ell}/2$, with $c$ being the speed of light. The vector containing all the distances of interest is defined as $\mathbf{d} = \left[d_0, d_1, \dots, d_{L-1} \right]^T$, with $L$ being the number of paths in the \ac{cir}.
If multiple \ac{cir} estimations are performed over a single packet, using different \acp{bp}, the reflections from the environments are amplified differently. This is due to the different \ac{bp} shapes, as each \ac{bp} steers the transmission signal towards a specific direction (beam steering). In addition, the \ac{cir} estimation is repeated for each packet $k$, which can be seen as sampling the \ac{cir} in time, with sampling period corresponding to the inter-packet transmission time $T_c$. The expression of the $\ell$-th \ac{cir} component, having delay $\tau_{\ell}$, obtained using beam-pattern $p$ at time (packet) $k$ is
\begin{equation}\label{eq:cir}
    h_{\ell, p}(k) = a_{\ell, p}(k) e^{j\phi_{\ell}(k)},
\end{equation}
where $a_{\ell, p}(k)$ and $\phi_{\ell}(k)$ are the complex gain of path $\ell$ at time $k$ and its phase, respectively. The path gain depends on the contribution of the \ac{bp} used for the transmission and on the reflectivity of the target, whereas the phase depends on the delay $\tau_{\ell}$. Note that $h_{\ell, p}(k)$ is a \textit{time domain} quantity, depending on the propagation delay index $\ell$ and on the time. In \eq{eq:cir} we used index $k$ as a shorthand notation for the discrete time instants $kT_c = 0, T_c, 2T_c,\dots $.

\subsection{People localization and tracking}
\label{sec:people-loc}

RAPID leverages the \ac{cir} estimates collected over time to continuously perform localization and tracking. The process develops in four steps: \textit{(i)} background subtraction, to remove the reflected paths due to static objects, \textit{(ii)} estimation of the subjects' distances, \textit{(iii)} estimation of the angular positions of the subjects with respect to the device, and \textit{(iv)} joint processing of distance and angle information using a Kalman filter to track each person's trajectory across time.

RAPID computes estimates at different rates, according to the specific resolution that is required by each task. 
Localization and tracking information are updated by RAPID every $\Delta t > T_c$ seconds, where index $t$ denotes the localization/tracking time-steps, whereas \ac{har} and identification require \ac{cir} readings at a rate $1/T_c$. The choice of setting $\Delta t > T_c$ stems from the fact that performing localization and tracking for every transmitted packet is unnecessary, as the packet transmission rate $1/T_c$ is much larger than the speed of human motion. This allows for additional flexibility in the selection of the type of \ac{bp}s that are used for each packet: as we explain shortly below in \secref{sec:11ay} and \secref{sec:imp}, we can modulate how many TRN units are included in a packet according to the type of sensing function that is being performed, i.e., localization/tracking versus activity/identity recognition. 


\subsubsection{Background subtraction}
\label{sec:back-sub}

To infer the positions of the subjects it is key to remove the reflections due to static (background) objects, as these typically have a much higher intensity than those generated by humans and may impact the localization accuracy. The background-related \ac{cir} is estimated by computing the time average of the \ac{cir} amplitude within a window of $K_{\rm static}$ samples, as static reflections are constant across time, 
\begin{equation}
    \bar{h}_{\ell, p} = \frac{1}{K_{\rm static}}\sum_{k=0}^{K_{\rm static}-1} |h_{\ell, p} (k)|.
\end{equation}
Then, the foreground \ac{cir} amplitude component is obtained as $|\tilde{h}_{\ell, p}(t)| = \max \left( |h_{\ell, p}(t)| - \bar{h}_{\ell, p}, 0 \right)$, i.e., removing the amplitude of the static paths and setting to zero the amplitude of those paths that would be present in the reference background \ac{cir}, but that are shielded by the presence of a person. We remark that, through different \acp{bp}, we perform beam steering at the transmitter. Hence, the peaks in $|\tilde{h}_{\ell, p}|$ correspond to the strongest propagation paths, as seen at the receiver when beam-pattern $p$ is used at the TX side. Changing the \ac{bp} $p$ allows scanning the environment by varying the transmission angle and, in turn, sweeping the whole field of view. We use this to infer the distance and the angular position of each individual, as described next.

\subsubsection{Distance estimation}
\label{sec:dist-est}

The distance of each subject is obtained by applying a threshold on $|\tilde{h}_{\ell, p}|$ (the time index is omitted for better readability), selecting the strongest paths across all the used \ac{bp}s. First, for each reflected path $\ell$, we consider vector 
\begin{equation}
\mathbf{h}_{\ell} = \left[ |\tilde{h}_{\ell, 0}|, |\tilde{h}_{\ell, 1}|, \dots, |\tilde{h}_{\ell, {N_p-1}}|\right]^T,
\end{equation}
containing the \ac{cir} values of path $\ell$ for each of the $N_p$ \ac{bp}s that are used at the transmitter. We collect the $L_2$-norms of $\mathbf{h}_{\ell}$, with $\ell=0, 1, \dots, L-1$,
 obtaining a new vector $\mathbf{h}$, as
 \begin{equation}
     \mathbf{h} = \left[ ||\mathbf{h}_{0}||_2, ||\mathbf{h}_{1}||_2, \dots, ||\mathbf{h}_{L-1}||_2\right]^T,
 \end{equation}
containing the strengths of each path at the receiver. We locate the local maxima in $\mathbf{h}$, denoting them by $h'_0, h'_1, \dots, h'_{\rm n_{peaks}-1}$. Hence, we discard those peaks with amplitude smaller than a dynamic threshold $A_{\rm th}$ computed from the maximum and average power of the paths in the current \ac{cir}. We introduce the following coefficients $\alpha_{\max}$, $\alpha_{\rm mean}$, and compute the threshold value $A_{\rm th}$, as
\begin{equation}\label{eq:threshold}
    A_{\rm th} = \max\left\{ \alpha_{\max} \cdot \max_{i} h'_i, \alpha_{\rm mean} \cdot \bar{h}'\right\},
\end{equation}
with $\bar{h}' = \sum_{i} h'_i / n_{\rm peaks}$.
A thorough evaluation of suitable values for $\alpha_{\max}$ and $\alpha_{\rm mean}$ is provided in \secref{sec:det-fa-results}.
With \eq{eq:threshold} the threshold is computed dynamically, proportionally to the maximum between the average and the maximum value of the \ac{cir}.
The peaks that exceed the threshold are selected as candidate targets of interest and used for the subsequent \ac{aoa} estimation. Denoting by $\ell_1, \ell_2, \dots, \ell_{N_s}$ the indices of the selected (candidate) paths ($0 \leq \ell_j \leq L-1$), the corresponding distances are obtained as $d_{\ell_j}=c\tau_{\ell_j}/2$.

\subsubsection{Angular position estimation}
\label{sec:aoa-est}
The following procedure is applied to each of the $N_s$ candidate paths. Let vector $\mathbf{s}_{\ell_j} \in \mathbb{R}^{N_p}$ contain the squared \ac{cir} amplitudes from one of such paths, $\ell_j$, for all used beam patterns, i.e., $\mathbf{s}_{\ell_j} = \left[|\tilde{h}_{\ell_j, 0}|^2, |\tilde{h}_{\ell_j, 1}|^2, \dots, |\tilde{h}_{\ell_j, N_{p}-1}|^2 \right]^T$. $\mathbf{s}_{\ell_j}$ is normalized by dividing it by its $L_2$-norm $||\mathbf{s}_{\ell_j}||_2$, then a correlation measure is used to estimate the angular position of the target by exploiting the gains of each beam pattern along the azimuth angular \ac{fov} $\theta$. Specifically, denoting by $g_p(\theta) \in [0,1]$ the normalized gain of beam pattern $p$ along direction $\theta$ (see \fig{fig:bp_shapes}), the angular position for candidate path $\ell_j$ is estimated as 
\begin{equation}\label{eq:aoa-corr}
    \theta_{\ell_j} = \argmax_{\theta} \sum_{p =0}^{N_p - 1} g_p(\theta) \frac{|\tilde{h}_{\ell_j, p}|^2}{||\mathbf{s}_{\ell_j}||_2^2}.
\end{equation}
The rationale behind \eq{eq:aoa-corr} is that if $|\tilde{h}_{\ell_j, p}|$ originates from the signal reflected off a subject, the corresponding angular direction is the one leading to the highest correlation between the \ac{cir} squared amplitude and the set of beam pattern gains. This is because each \ac{bp} amplifies path $\ell_j$ differently, depending on the beam pointing direction.

Upon obtaining the distance and the angle estimates, an Extended Kalman filter is utilized to track the subjects' positions over time.

\subsubsection{People tracking - extended Kalman filter}
\label{sec:ekf}

After the localization step, the candidate positions of the subjects are known in polar coordinates, and constitute our \textit{observations} of the positions of the subjects, which we denote by $\mathrm{\mathbf{z}}_t^j = [d_{\ell_j}, \theta_{\ell_j}]^T, \forall j=1,2,\dots,N_s$.
We employ an \ac{ekf} \cite{ribeiro2004kalman} to track the physical position of each individual in the Cartesian space. Specifically, we define the true state of subject $j$ at time $t$ as vector $\mathrm{\mathbf{x}}^j_t = \left[x^j_t, y^j_t, \dot{x}^j_t, \dot{y}^j_t\right]^T$, containing the coordinates along the $x-y$ horizontal plane and the movement velocity components along the same axes. We approximate the motion of the subjects with a constant velocity (CV) model \cite{schubert2008comparison}.
As the observations $\mathrm{\mathbf{z}}_t^j$ become available, we apply the predict and update steps of the \ac{ekf} to follow the movement trajectories of the subjects \cite{ribeiro2004kalman}.
The association between the observations from time $t+1$ and the states from time $t$ is done using the nearest-neighbors joint probabilistic data association algorithm (NN-JPDA) \cite{shalom2009probabilistic}.

Using the \ac{ekf} estimates $\mathrm{\hat{\mathbf{x}}}^j_t$ of each person's state across subsequent time steps allows retrieving the path and the \ac{bp}s in the \ac{cir} which contain his/her \ac{md} signature.

\subsection{micro-Doppler extraction}
\label{sec:mu-dopp}

\subsubsection{\ac{cir} phase model}
\label{sec:cir-phase-model}
The \ac{cir} model in \eq{eq:cir} is here expanded and related to radar theory \cite{patole2017automotive}. 
Using a typical radar terminology, we refer to the \ac{cir} samples $\ell = 0, 1, \dots, L-1$ as the \textit{fast-time} sampling dimension, as they are obtained at the highest available sampling rate. 
The \ac{cir} samples collected across different packets are instead referred to as the \textit{slow-time} samples, indicized by variable $k$ as in \secref{sec:cir-model}.

Next, we consider a moving object within the monitored indoor space; the transmitted signal is reflected off the object and the corresponding contribution is retrieved at the receiver in the $\ell$-th path of the \ac{cir}. To extract the \ac{md} effect caused by the movement of this object, we analyze the phase of the $\ell$-th path across time. The time-dependent phase term in \eq{eq:cir} can be expressed as follows
\begin{equation}\label{eq:cir-phase}
    \phi_{\ell}(k) = -2\pi f_o \frac{2\left(d_{\ell} - v_{\ell} k T_c\right)}{c} = -2\pi f_o \bar{\tau}_{\ell} + 4 \pi f_o \frac{v_{\ell}}{c} k T_c.
\end{equation}
Here, $\bar{\tau}_{\ell}$ is the delay of the $\ell$-th path due to the distance of the corresponding reflector from the device.
$v_{\ell}$ is the radial velocity of the reflector with respect to the device, which is assumed to be \textit{slowly} time-varying, i.e, we can consider it constant during a \ac{md} spectrum processing interval (see \secref{sec:md-spec}). From \eq{eq:cir-phase} it can be seen that the velocity of the object at distance $d_{\ell}$, if greater than zero, modulates \ac{cir} phase across the slow time dimension. Following a common convention \cite{patole2017automotive}, in this work objects moving away from the transmitter (\ac{ap}) have \textit{positive} velocity, while incoming objects have \textit{negative} velocity.

The human body contains multiple moving parts that have different velocities and follow different trajectories. Thanks to the small wavelength of \acp{mmwave}, in the \ac{md} we can observe these different contributions via \ac{tf} analysis, as detailed in the next \secref{sec:md-spec}. 

\subsubsection{micro-Doppler spectrum}
\label{sec:md-spec}

Human movement causes a frequency modulation on the reflected signal due to the small-scale Doppler effect produced by the different body parts. Using \ac{tf} analysis of the received signal, it is possible to distinguish between different actions performed by a person or identify the individual based on his/her way of walking (\textit{gait}) \cite{seifert2019toward, vandersmissen2018indoor}. 
\ac{mmwave} radios are particularly suited for this, as their frequencies are sensitive to the \ac{md} effect due to their small wavelengths.

From \eq{eq:cir-phase}, the \ac{md} effect of human movement can be extracted from subsequent estimates of the \ac{cir}, computed every $T_c$ seconds. Specifically, one can compute the short-time Fourier transform (STFT) of $h_{\ell, p}(k)$, across slow-time, for each path $\ell$ and each beam pattern $p$ as
\begin{equation}\label{eq:stft}
    H_{\ell, p}(n, i) =\sum_{m=0}^{M-1} h_{\ell,p}(m + n\sigma) w(m) e^{ -j2\pi\frac{im}{M}},
\end{equation}
where $n$ is the time index, $i=0, 1, \dots, N_D-1$ is the frequency index, $M$ is the (fixed) window length, $w$ is a Hann window of dimension $M$ and $\sigma$ is the time granularity of the STFT.
The power spectrum of $h_{\ell, p}(k)$, computed as $\mu_{\ell, p} (n, i) = |H_{\ell, p}(n, i)|^2$, contains information on the phase modulation due to the velocity $v_{\ell}$, and can be used to analyze its evolution across subsequent windows. 

\eq{eq:stft} can not be used directly to extract the \ac{md} signature of a moving human in our setup, as it refers to a single fast time bin (a single path in the \ac{cir}) and a single \ac{bp}, while people can be located in different positions across time. In addition, it would be inefficient to compute the STFT for all the paths and all the \ac{bp}s. Instead, the computation should only be performed for those physical locations where a person is detected.
In the following, we leverage the localization and tracking process described in \secref{sec:people-loc} to only extract the \ac{cir} portions that contain useful \ac{md} information.

\subsubsection{\ac{md} separation}
\label{sec:md-sep}

Assume that we want to extract the \ac{md} of a person that was detected and located by the previous algorithms at a certain distance and angle with respect to the device. Hence, we extract the \ac{cir} samples from the most useful \ac{bp}, i.e., the one that points in the direction of the person and that, in turn, emphasizes the most the reflection from this target.

From the estimated state of this person (\secref{sec:ekf}), their angular position is obtained as $\hat{\theta}=\arctan\left(\hat{y}_t / \hat{x}_t\right)$ and their distance from the device, as $\hat{R}=\sqrt{\hat{x}_t^2 + \hat{y}_t^2}$. The \ac{bp} approximately pointing in the direction of this person, denoted by $p^*$, is thus selected as the \ac{bp} having the highest gain along $\hat{\theta}$, that is
\begin{equation}\label{eq:bp-person}
p^* = \argmax_{p} g_p(\hat{\theta}).
\end{equation}
Moreover, due to the high ranging accuracy of \acp{mmwave}, humans typically produce reflections that influence more than a single \ac{cir} path. The \ac{cir} paths of interest are those that correspond to a neighbourhood of $\hat{R}$. In our analysis, we take the size of this neighborhood constant across all subjects, denoting it by $Q$. Specifically, we first select the path $\ell^*$ that best matches the subject's distance $\hat{R}$
\begin{equation}
    \ell^* = \argmin_{\ell} |d_{\ell} - \hat{R}|.
\end{equation}
Then, from the original complex-valued \ac{cir}, we extract a window containing $Q+1$ samples along the fast-time dimension, centered on $\ell^*$, and use \eq{eq:stft} to compute the \ac{md} spectrum components for our target at time $n$ as  
\begin{equation}
    \mu_i(n) =\sum_{\ell=\ell^* - Q/2 }^{\ell^* + Q/2} \left|H_{\ell, p^*}(n, i)\right|^2, \quad i=0, 1, \dots, N_D-1.
\end{equation}
These \ac{md} spectrum components are collected through vector $\boldsymbol{\mu}(n) = \left[\mu_0(n), \mu_1(n), \dots, \mu_{N_D - 1}(n)\right]^T$.

To capture the human movement evolution across time, we compute the \ac{md} vectors for a window of $N_{\mu{\rm D}}$ subsequent time-steps and concatenate them into a spectrogram representing the \ac{md} signature of the target up to time $n$, as
\begin{equation}\label{eq:mud-spec}
    \boldsymbol{\Upsilon}_n = \left[\boldsymbol{\mu}(n - N_{\mu{\rm D}} +1), \boldsymbol{\mu}(n - N_{\mu{\rm D}} +2), \dots, \boldsymbol{\mu}(n) \right].
\end{equation}
The procedure described in this section is repeated for all the detected subjects. 

\subsubsection{Human \ac{md} range and resolution} \label{sec:md-res}

During everyday movement, the limbs of a person usually have velocities of up to $3-4$~m/s \cite{vandersmissen2018indoor, seifert2019toward}. To fully capture the \ac{md} signature of the subjects, we must ensure that our systems achieves a sufficient resolution. Recalling \eq{eq:cir-phase}, we know that the Doppler frequency shift induced by a moving object on the $\ell$-th path is $f_{\ell}^{\rm D} = 2 f_o v_{\ell}/c$.
Using \ac{tf} analysis to estimate the Doppler spectrum as in \eq{eq:stft}, the resolution that can be obtained on the Doppler frequency is $\Delta f^{\rm D} = 1/(MT_c)$. The maximum measurable Doppler frequency is instead $f^{\rm D}_{\rm max} = 1/(2T_c)$. These quantities can be mapped onto the velocity estimate resolution and the maximum measurable velocity as
\begin{equation}\label{eq:dvel-res-range}
    \Delta v = \frac{c}{2f_o  M T_c}, \quad v_{\rm max} = \frac{c}{4f_o T_c}.
\end{equation}
Given that we sample the \ac{cir} on a per-packet basis, to capture the \ac{md} effect of human motion we must ensure that the time $T_c$ between the packets used in the \ac{md} estimation allows capturing the range of velocities of interest. See also \secref{sec:ex-res} for the chosen values of $M$ and $T_c$.

\subsection{Activity recognition and person identification}
\label{sec:har-id}
\begin{figure*}[t!]
	\begin{center}   
		\centering
		\subcaptionbox*{Walking. \label{fig:walking_5sub}}[4cm]{\includegraphics[width=4cm]{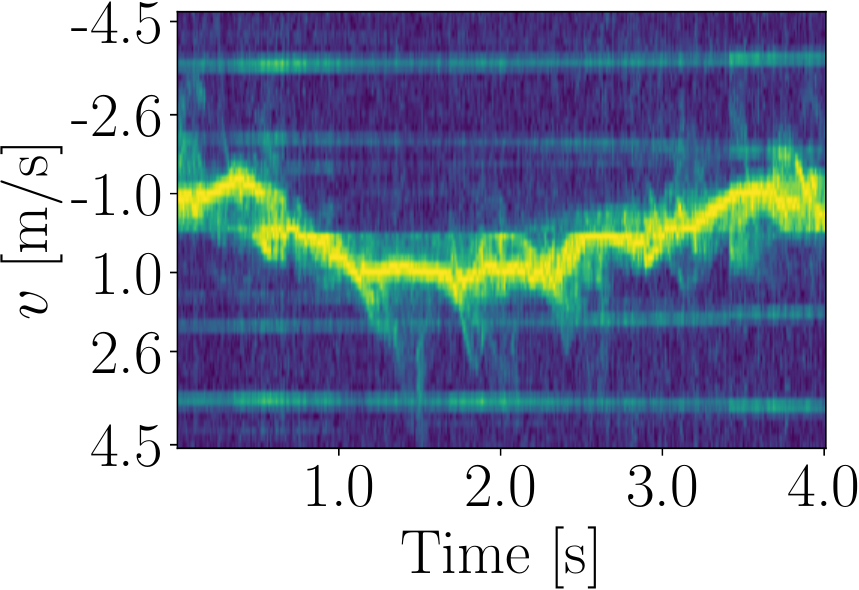}}
		\subcaptionbox*{Running. \label{fig:running_5sub}}[4cm]{\includegraphics[width=4cm]{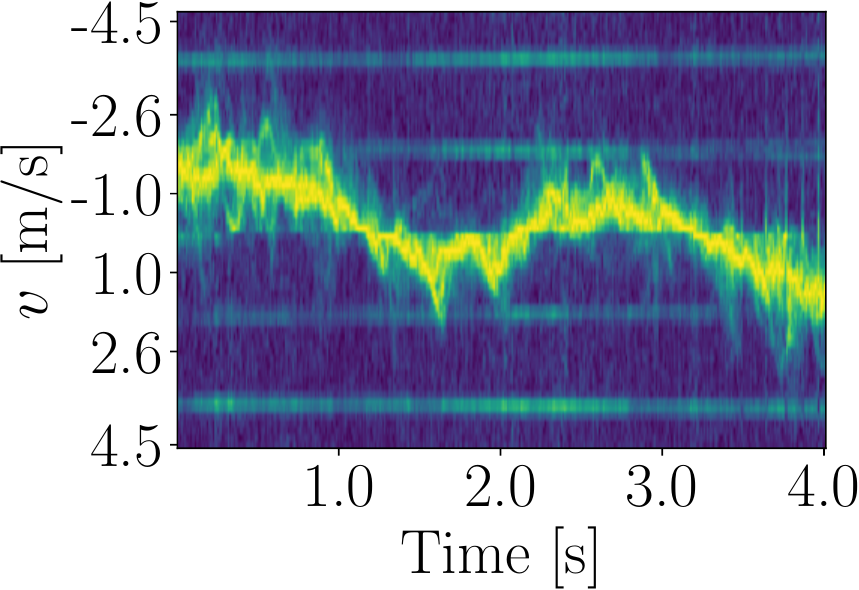}}
		\subcaptionbox*{Sitting down. \label{fig:sitting_5sub}}[4cm]{\includegraphics[width=4cm]{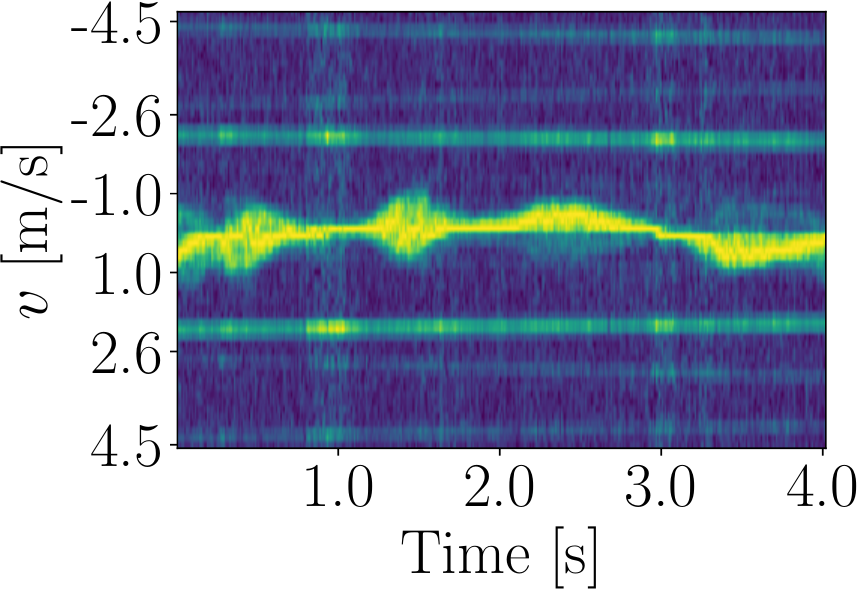}}
		\subcaptionbox*{Waving hands.\label{fig:hands_5sub}}[4cm]{\includegraphics[width=4cm]{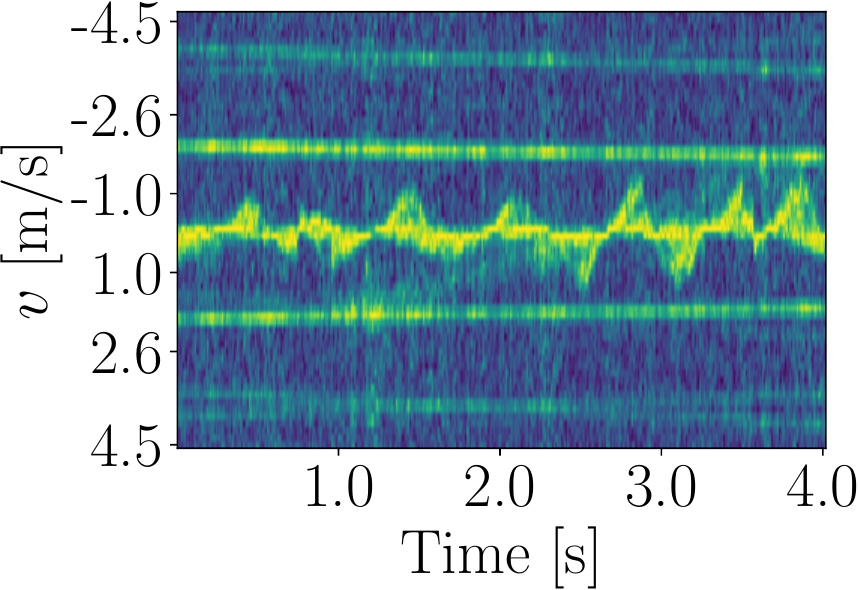}}
		\caption{Example $4$~s long \ac{md} spectrograms obtained by RAPID from $4$ subjects. The yellow and blue colors respectively represent high and low power in the corresponding Doppler velocity bins ($y$ axis).}
		\label{fig:md-act}
	\end{center}
 \vspace{-0.4cm}
\end{figure*}

The \ac{md} signature, obtained as in \eq{eq:mud-spec}, contains information about the type of movement performed by the person.

To perform \ac{har} and person identification, we use a deep neural network to classify each spectrogram. Specifically, once the \ac{md} signatures of each person have been separated, RAPID performs the following tasks: \textit{(i)} it classifies the activity carried out by the subject into \textit{walking} (A$0$), \textit{running} (A$1$), \textit{sitting down} (A$2$), \textit{waving hands} (A$3$) and \textit{standing still} (A$4$) and \textit{(ii)} it recognizes the subject's identity during a walking phase, among a known set of individuals, denoted by S$0$, S$1$, etc. In \fig{fig:md-act}, we show \ac{md} signature examples for activities A$0-3$, concurrently performed by $4$ subjects within the same environment. 

As human \ac{md} is highly variable across different subjects, and we seek robustness to different environment conditions and noise, we employ deep learning to classify the \ac{md} signatures.
Referring to a single subject, the \ac{md} spectrum $\boldsymbol{\Upsilon}_n$ is represented as an image and processed by two separate \acp{cnn} for \ac{har} and person identification, respectively. The two classifiers share the same architecture, as shown in \fig{fig:dl-diagram}, but are trained separately and have different weights as they perform different tasks. As the subjects are continuously tracked over time, we adopt a sliding window approach, selecting \ac{md} spectrograms with $N_{\mu{\rm D}}$ \ac{md} spectrum samples for each window (matrix $\boldsymbol{\Upsilon}_n$). Subsequent windows partially overlap to increase the reactiveness of RAPID in obtaining predictions. Both \acp{cnn} are trained to extract features from the \ac{md} spectrograms and to classify the activity performed by or identity of the person, by learning a function $\mathcal{F}(\cdot)$ that maps a \ac{md} window, $\boldsymbol{\Upsilon}_n$, of size $N_{D} \times N_{\mu{\rm D}}$, onto a vector $\mathbf{c}_n$ containing the \ac{har} (identification) class probabilities, i.e., $\mathbf{c}_n = \mathcal{F}(\boldsymbol{\Upsilon}_n)$. The dimension of the final probability vector $\mathbf{c}_n$ is different in case of \ac{har} or identification depending on the dimension of the classification problem. The second \ac{cnn}, used for person identification, is only trained on walking spectrograms, as human gait is well known to be a soft biometric identifier \cite{nambiar2019gait}. Hence, during the system operation, the identification classifier is only applied on the input \ac{md} spectrogram when the activity is classified as ``walking'' by the \ac{har} classifier, see \fig{fig:dl-diagram}.

\subsubsection{\ac{md} spectrogram pre-processing}\label{sec:md-preproc}

Prior to feeding it to the \ac{cnn} classifier, the \ac{md} spectrogram is pre-processed by removing the contributions from static reflections and normalizing it.

\smallsection{Static reflection removal} A customary step when processing human \ac{md} signatures is the removal of static reflections, which appear as a strong power peak around the $0$~m/s velocity bin. This can be done by either applying a high-pass filter to the signal or, if deep learning methods are used for classification, by directly removing the Doppler bins containing unwanted contributions, as done in \cite{vandersmissen2018indoor,pegoraro2021multiperson}. We adopt the latter method to remove the Doppler bins corresponding to the velocities in the interval $[-0.28, 0.28]$~m/s, as they contain very low, non-informative velocities.

\smallsection{Normalization} To compensate for differences in the strength of the reflections when subjects are far from the \ac{ap}s, we normalize each column of $\boldsymbol{\Upsilon}_n$, $\boldsymbol{\mu}(j), j=0, 1, \dots, N_{\mu{\rm D}}-1$ in the range $[0, 1]$.


\subsubsection{Deep learning classifier}\label{sec:dl-classifier}

\begin{figure*}
    \centering
    \includegraphics[width=16cm]{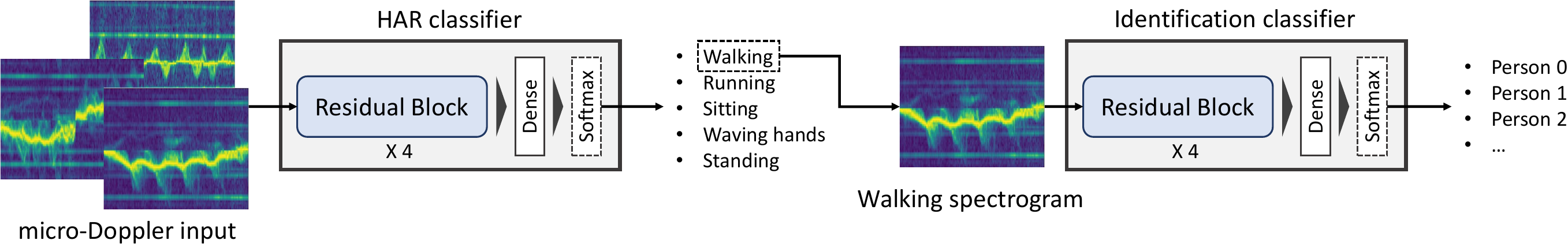}
    \caption{Block diagram of the \ac{cnn} classifiers used by RAPID for \ac{har} and person identification.}
    \label{fig:dl-diagram}
\end{figure*}

We use the same \ac{cnn} architecture, based on deep residual networks \cite{he2016deep}, for \ac{har} and person identification, with the only difference being the dimension of the last classification layer. This network consists of $4$ consecutive residual blocks. Each residual block has two convolutional layers \cite{goodfellow2016deep}, the first of which includes a down-sampling by a factor of $2$ (\textit{stride}). Each convolution is followed by an \ac{elu} activation function \cite{clevert2015fast} and batch normalization \cite{ioffe2015batch}. The output of the convolution is summed to the input (\textit{skip connection}) and passed through another \ac{elu} activation and batch normalization. The $4$ residual blocks use $8$, $16$, $32$ and $64$ filters, respectively, all having a kernel of size $3 \times 3$. After the last residual block, we apply Dropout \cite{srivastava2014dropout} with a ratio of $0.5$, and a fully-connected (or \textit{dense}) layer with $64$ units, then, a second Dropout operation with ratio of $0.2$. Finally, the classification probabilities for \ac{har} or person identification are computed via a Softmax activation function \cite{goodfellow2016deep}. The network architecture is shown in \fig{fig:dl-diagram}. 

\subsubsection{Combining multiple \ac{ap}s}\label{sec:dec-fus}

Using the different points of view provided by the different \ac{ap}s, RAPID can improve its \ac{har} and person identification performance. Assume that a person is independently detected and tracked by $2$ or more \ac{ap}s concurrently. A slightly different \ac{md} signature of the person is obtained by each \ac{ap}, according to the angular position and the distance of the device with respect to the person. At each time instant $n$, we adopt a simple decision fusion scheme including the following steps: \textit{(i)} if a single \ac{ap} detects the person, the decision made by the classifier on the corresponding \ac{md} signature is used, i.e., $\argmax_j c_{n, j}$, where $c_{n,j}$ is element $j$ of vector $\mathbf{c}_n$, \textit{(ii)} if multiple \ac{ap}s detect the person, denote by $\mathbf{c}_n^a$ the probability vector predicted by \ac{ap} $a$. The final decision is made by the \ac{ap} that is most confident about its classification, i.e., the one that assigns the highest probability to the predicted class: $\argmax_j \left\{\max_a c_{n, j}^a \right\}$.

\section{Enabling sensing in IEEE 802.11ay} \label{sec:11ay}

The high bandwidth of IEEE 802.11ay \cite{802.11ay} not only provides high data throughput but also offers excellent accuracy for sensing applications. RAPID is able to extract highly accurate range, angle and \ac{md} information from \ac{cir} measurements. For this, we take advantage of the beam training and beam tracking mechanisms of IEEE 802.11ay systems. 

Range and angle information are extracted from the \ac{cir} obtained via the \acp{cef} of standard beacon frames that are frequently sent by the \ac{ap} or the beam training frames sent during a \ac{sls}. The \ac{sls} is a two-step procedure: first, one device sends training frames using the available antenna configurations, while the second device listens using a quasi omnidirectional \ac{bp}. Then, the devices exchange their roles to train the other device. After sending feedback, the devices can select the \textit{best} combination of \ac{bp} on both sides of the link. IEEE 802.11ay also introduces the concept of \textit{in-packet} beam tracking \cite{Knightly_ICOMM2017}, where different antenna configurations can be tested within a single packet, allowing for much quicker \ac{bp} changes. This is done by appending a TRN field to the packet as shown in \fig{fig:11ay_packet}. A TRN field is composed of multiple (variable) TRN units formed by $6$ complementary Golay sequences of type a (``Ga'') and b (``Gb'') with length $128$~samples each:
%
\begin{figure}
    \centering
    \includegraphics[width=\linewidth]{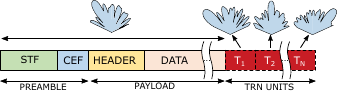}
    \caption{IEEE 802.11ay \textit{in-packet} training packet.}
    \label{fig:11ay_packet}
\end{figure}
\begin{equation} \label{eq:TRN}
\left\{ \text{+Ga}_{128} \text{; -Gb}_{128} \text{; +Ga}_{128} \text{; +Gb}_{128} \text{; +Ga}_{128} \text{; -Gb}_{128} \right\}.
\end{equation}
The excellent autocorrelation properties of the complementary Golay sequences and the availability fast hardware structures for the correlation \cite{liu_digital_2013} make them ideal for \ac{cir} estimation \cite{Lacruz_MOBISYS2021}. The high bandwidth (1.76~GHz) results in a range resolution of $\sim 8.5$~cm directly from the \ac{cir} estimate.
%
Considering the different \ac{bp} shapes used during beam training, possible targets located in the \ac{fov} of the devices are \textit{illuminated} by the respective \ac{bp}s that focus energy in that direction and they appear as multi-path components in the \ac{cir} (Fig. \ref{fig:cir}). Furthermore, we take advantage of the different amplification factors in the multi-path components (given by the different \ac{bp}s) to estimate the angular positions of the subjects. For this purpose, we apply the correlation based approach explained in \secref{sec:aoa-est} to the different channel multi-path components in the channel. 
Considering the common speed of human motion, carrying out beaconing or a beam training procedure every, e.g., $100$~ms allows accurately locating human targets in the \ac{fov} of the \acp{ap}. Note that, as we show in \secref{sec:ex-res}, a \textit{full} beam training, that scans all the available \acp{bp}, is in fact not needed and we may use a much smaller subset of \acp{bp}.

Extracting \ac{md} signatures from the \ac{cir} requires fine-grained frequency resolution, as detailed in \secref{sec:md-res}. This cannot be achieved with the \ac{cir} estimates obtained from beacons or beam training packets only, as sampling the \ac{cir} with $T_c = 100$~ms would lead to an insufficient maximum measurable Doppler velocity of $6.25\cdot10^{-3}$~m/s (see \eq{eq:dvel-res-range}). We address this by exploiting the \textit{beam tracking} procedure defined in the standard \cite{802.11ay}. It allows to add a configurable number of TRN units to data packets to test different \ac{bp} configurations to \textit{quickly} correct possible misalignment without requiring a full beam training procedure. 

After identifying the subjects' ranges and angles using beam training packets, we include a TRN field in \textit{data packets} with a sufficient number of TRN units to illuminate all the subjects in the scene; each TRN unit uses a suitable \ac{bp} that specifically points in the direction of a person. This steers the energy of the transmitted signal so as to best capture the \ac{md} signatures of the subjects, while maintaining low additional overhead for the data packets. Considering that data packets are sent much more frequently than beam training packets, our approach can sample the \ac{cir} with a sufficiently low $T_c$ to capture the desired range of frequencies for human movement analysis. 

\section{Implementation} \label{sec:imp}

The available \ac{mmwave} \ac{cots} devices support IEEE 802.11ad and offer very limited access to physical layer information \cite{steinmetzer2017compressive}. To the best of the authors' knowledge, there are no \ac{cots} solutions for the new IEEE 802.11ay standard available yet. To address the lack of hardware, we turn a \ac{mmwave} \ac{sdr} system into a \ac{jcr} experimentation platform. Here we cover the design decisions made to implement RAPID on such platform. 

\subsection{Hardware components}

As a baseline to implement a RAPID \ac{ap}, we use the mm-FLEX experimental platform \cite{Lacruz_MOBISYS2020}. This open platform is composed of a baseband processor including a Xilinx Kintex Ultrascale FPGA plus high-speed AD/DA converters and DDR memory banks. Besides, it is connected through a PCIe interface to a Core i7 processor card co-located within the same hosting chassis. The latter implements configuration and control tasks for the whole system. 

The baseband processor is configured to fulfill the bandwidth requirements of IEEE 802.11ad/ay standards ($1.76$~GHz), using a sampling frequency of $3.52$~GSPS for both AD/DA converters, with $2$ samples per symbol.

\begin{figure*}
    \centering
    \begin{minipage}{0.76\linewidth}
        \begin{minipage}{0.5\linewidth}
        \begin{subfigure}[b]{\linewidth}
            \centering
            \includegraphics[width=\linewidth]{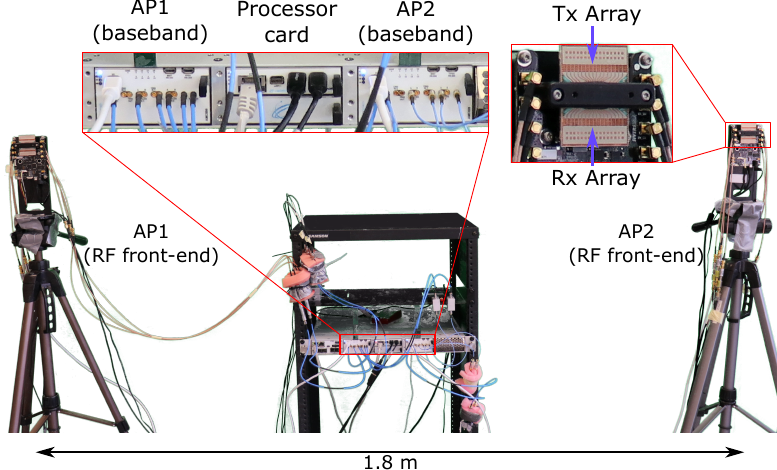}
            \vspace{0.1cm}
            \caption{Two RAPID \ac{ap}s deployment}
            \label{fig:testbed}
        \end{subfigure}
        \end{minipage}
        \hspace{0.2cm}
        \begin{minipage}{0.46\linewidth}
            \begin{subfigure}[b]{0.48\linewidth}
                \centering
                \input{FIGURES/BP_SHAPES.tikz}
                \vspace{-0.6cm}
                \caption{Beam pattern shapes}
                \label{fig:bp_shapes}
            \end{subfigure}
            \begin{subfigure}[b]{0.48\linewidth}
                \centering
                \input{FIGURES/CFO.tikz}
                \vspace{-0.6cm}
                \caption{Measured CFO}
                \label{fig:CFO}
            \end{subfigure} 
            \begin{subfigure}[b]{\linewidth}
                \centering
                \input{FIGURES/CIR_PLOT.tikz}
                \vspace{-0.25cm}
                \caption{Example of CIR measurement}
                \label{fig:cir}
            \end{subfigure}
        \end{minipage}
        \vspace{-0.2cm}
        \caption{{RAPID implementation}}
    \end{minipage}
    \begin{minipage}{0.21\linewidth}
        \centering
        \includegraphics[width=0.95\linewidth]{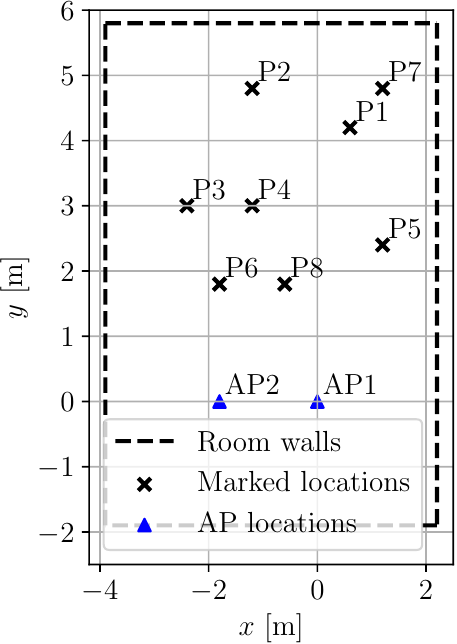}
        \caption{Schematic representation of E$1$.}
        \label{fig:scenario}
    \end{minipage}
\end{figure*}

The RF front-end includes a $60$~GHz up/down converter and phased antenna arrays from Sivers \cite{SIVERSIMA}. The device is able to operate on all the channels defined in the IEEE~802.11ad/ay standards \cite{802.11ad,802.11ay}. As shown in \fig{fig:testbed}, the device integrates two independent $16$-element linear antenna arrays, one used for transmission and one for reception. The codebook of \acp{bp} for both arrays can be freely configured.
The system is controlled in real-time using USB and SPI interfaces, as well as GPIO pulses for the quick \ac{bp} changes required for beam training and tracking.

\begin{figure}
    \centering
    \includegraphics[width=0.51\linewidth]{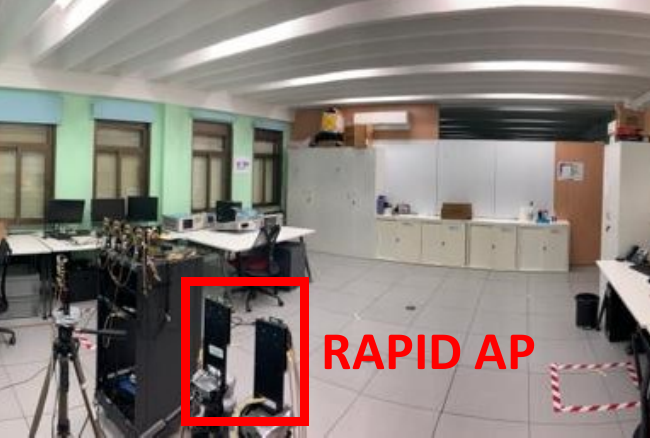}
    \includegraphics[width=0.45\linewidth]{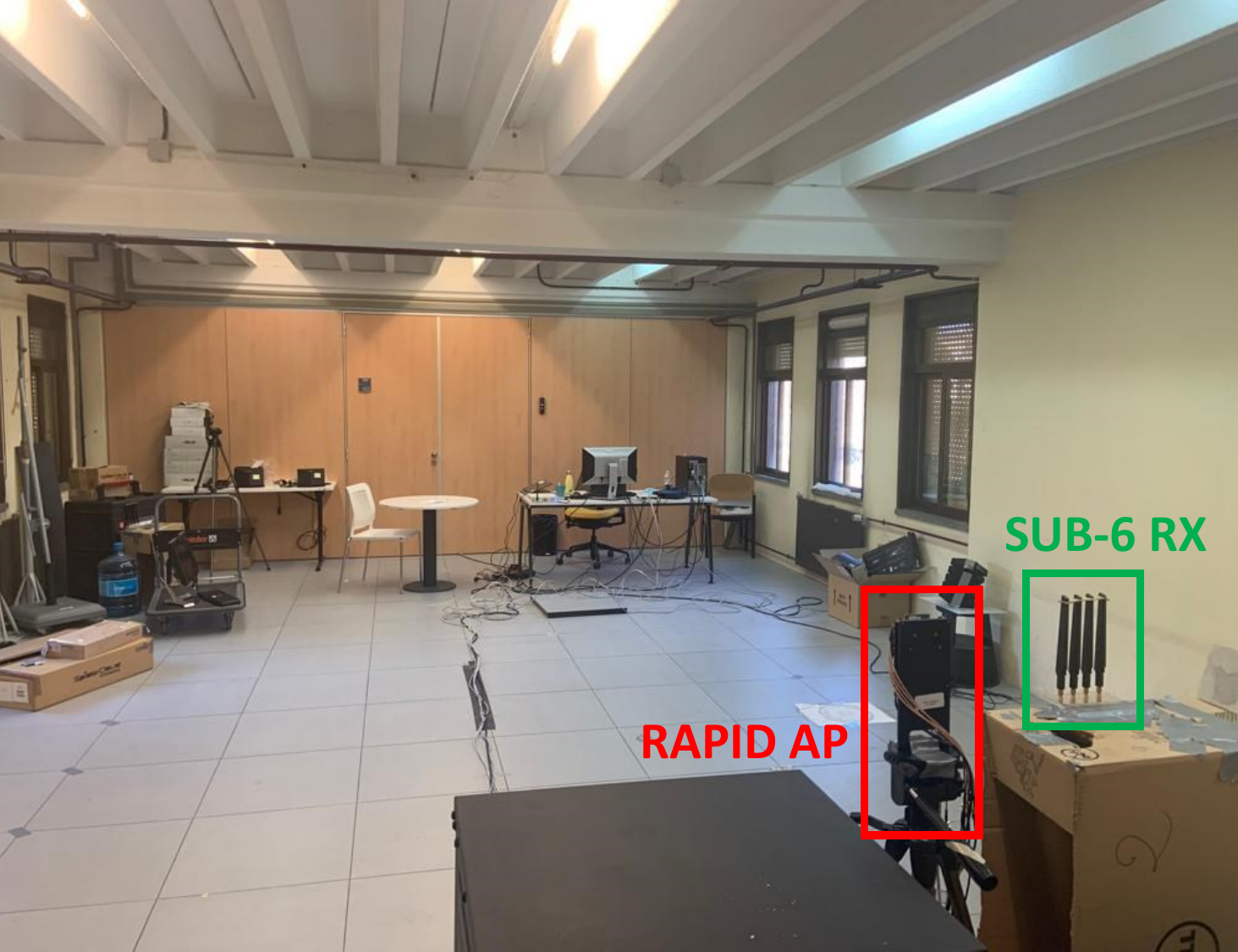}
    \caption{The two environments: E$1$ (left) and E$2$ (right).}
    \label{fig:env-pics}
\end{figure}

\subsection{Full-duplex operation}

To bring radar capabilities to the experimentation platform, it is necessary to support simultaneous operation of the TX and RX chains. This is achieved by concurrently enabling transmit and receive sub-systems in the RF front-end, and by enhancing the functionality of the baseband processor.

The $60$~GHz front-ends used in this work \cite{SIVERSIMA} are laboratory equipment designed for early stage proof-of-concept communication systems. The carrier frequency is generated from a $45$~MHz clock, which introduces significant \ac{cfo} and destroys the phase coherence between the \ac{cir} estimates obtained from consecutive packets. This would make the extraction of \ac{md} signatures infeasible with two independent co-located antennas.
Instead, by using both transmit and receive arrays from the same RF front-end (see \fig{fig:testbed}), up and down conversion sub-systems are fed by the same local oscillator which keeps \ac{cfo} levels in the range of $[-40, 40]$~Hz, as shown in \fig{fig:CFO}. Although transmit and receive arrays are directly next to each other, \emph{no complex analog or digital self-interference cancellation techniques are required}. Thanks to the directional \ac{bp}s and the robustness of the Golay Sequences of the TRN units, the system only requires some transmit power control to avoid saturating the receive antennas and down-conversion stages. 
In \fig{fig:cir}, we show the \ac{cir} measurements obtained from multiple \ac{bp}s within a packet, by marking the self interference path and the reflections from the test room, where the different amplitudes correspond to the different \ac{bp} shapes towards the direction of the reflectors. 

In the baseband processor, we implement a state-machine on the FPGA logic which controls the transmit and receive data-paths. Specifically, it handles the DDR memory that stores the transmit frames, performs multiple real-time antenna reconfigurations over the TRN field of the packet, triggers the DDR memory on the receive data-path, and sets the inter-frame spacing between multiple transmitted packets. 
While here we focus on an \ac{ap}-centric design, the same procedure can be applied to implement RAPID on any station in the network. 

Since our RAPID AP operates in a mono-static configuration, we perform \ac{cir} extraction without requiring the use of packet detection and synchronization circuits. To do this, it is important to ensure deterministic latency between the transmit and receive data-paths. Considering that transmit and receive data-paths have their own independent clock structure, we use clock-domain crossing techniques to send the state machine signals across transmit and receive domains. Besides, latencies in the DDR controllers are variable, which requires the use of FIFO queues at the output/input of the TX/RX DDRs.
Together, these solutions help to achieve the desired deterministic latency.


\subsection{Multi-AP system}

Since IEEE 802.11ay networks typically involve many \acp{ap} and dense deployments,
we extend the aforementioned testbed capabilities to handle multi-AP scenarios. To this end, we integrate a second baseband processor in the hosting chassis which is connected to an independent $60$~GHz front-end. Each \ac{ap} has their its clocking structure, i.e., \acp{ap} are not synchronized. 
Each \ac{ap} can be freely configured with its own parameters. For the sake of simplifying the system management, we use different communication channels ($58.32$ and $60.48$~GHz) for each \ac{rf} front-end, avoiding cross interference. It is worth mentioning that the channels can be freely configured, making it possible to operate the two RAPID \ac{ap}s so that they share the same frequency band, by implementing carrier sensing mechanisms.


 
\section{Experimental results} \label{sec:ex-res}

In this section, we discuss the results of our extensive measurement campaign.
Motivated by the discussion in \secref{sec:md-res} and \secref{sec:11ay}, for the \ac{md} estimation we consider data packets (with TRN fields) spaced by $T_c = 0.27$~ms. This allows capturing velocities in the range $[-4.62, 4.62]$~m/s and leads to a resolution of $\Delta v = 0.14$~m/s when using a window of $M=64$ samples in the DFT computation, see \eq{eq:dvel-res-range}. These values are comparable to the ones achieved with radar devices \cite{zhao2019mid, pegoraro2021multiperson, vandersmissen2018indoor}. Note that the even spacing of packets is just for convenience but is not a requirement, i.e., estimation can be done with random bursts of data packets with sufficiently small spacing. Moreover, we set to $Q=9$ the size of the fast time window used to capture the contribution of the subjects in the \ac{cir} (see \secref{sec:md-sep}). The EKF time-step duration is set to $\Delta t = 32 T_c$, which is also the time-granularity at which we obtain \ac{md} spectrum vectors.
To extract range and angle information, we use in-packet beam training frames with $12$ TRN units, using antenna beams covering a \ac{fov} range from $-45^\circ$ to $45^\circ$. With this configuration we achieve a mean accuracy of $2^\circ$ for the angular position of a person standing in the room. We verify that this allows tracking multiple subjects reliably and with low localization error, as detailed in the following. In order to implement the angle estimation method from \secref{sec:aoa-est}, we measured the \ac{bp} shapes from the codebook using a motorized pan-tilt platform. In \fig{fig:bp_shapes}, we show the $12$ \ac{bp}s we used to perform the experiments. 

\subsection{Experiment setup}


We test RAPID in two different rooms, as shown in \fig{fig:env-pics}. The two environments are research laboratories, denoted by E$1$, of dimensions $6.1 \times 7.7$~m and E$2$, of dimensions $6 \times 10.7$~m (E$2$), and containing whiteboards, windows, tables, computers and equipment, making them challenging multi-path environments with a number of potential reflectors. Most of our experiments, including the collection of the training data for the NN classifier, have been carried out in E$1$, while we used E$2$ to assess the robustness of the proposed method to unknown environments. For the tests involving multiple \acp{ap}, we deploy two RAPID \ac{ap}s as shown in \fig{fig:testbed} close to the wall, separated by $1.8$~m.

To test the localization and tracking capabilities of RAPID, we mark specific known positions across E$1$ to determine the ground truth location as shown in \fig{fig:scenario}, and perform our tests by having subjects move across these positions.
The markers are denoted by P$x$, with $x$ ranging from $1$ to $8$, while \ac{ap}s are represented as blue triangles. The room walls are represented with a black dashed line.

\subsection{Baseline experiments} \label{sec:prel-exp}

We first report the results obtained in two simple baseline experiments to verify the capability of RAPID to extract the \ac{md} signature of a moving person in an indoor scene.
Here, we only use \ac{ap}~$1$ and a single subject, performing different activities at different locations in E$1$.

\begin{figure}[t!]
    \centering
    \includegraphics[width=4cm]{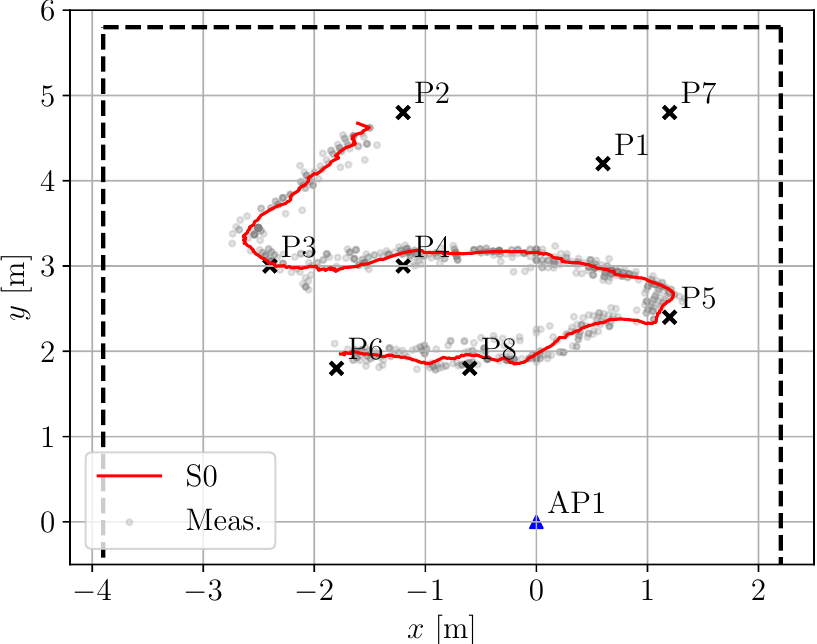}
    \includegraphics[width=4cm]{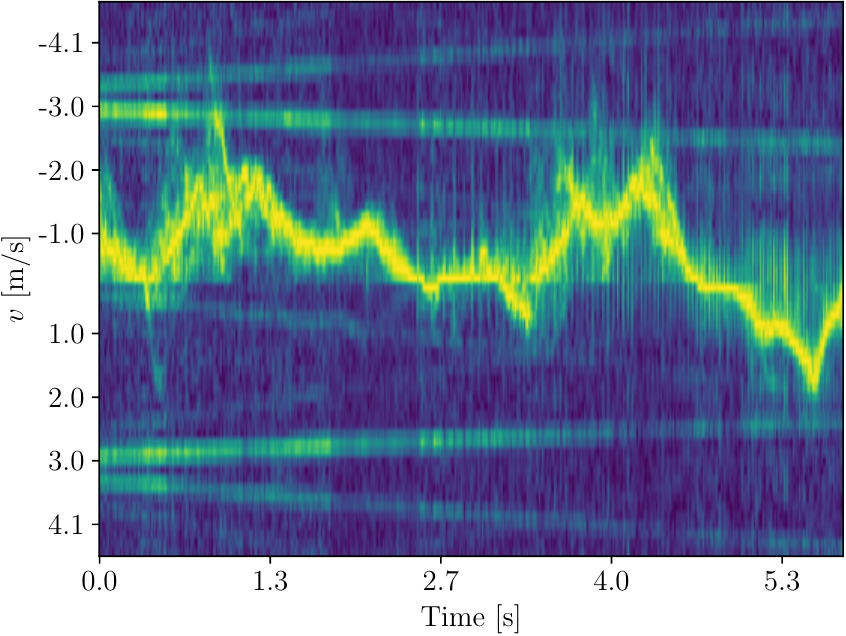}
    \caption{Subject walking trajectory (left) and a portion of the corresponding \ac{md} signature (right) extracted by RAPID.}
    \label{fig:walk_1sub}
\end{figure}

\fig{fig:walk_1sub} shows the EKF estimated trajectory of the subject walking along the trajectory P$2$-P$3$-P$4$-P$5$-P$8$-P$6$ together with the corresponding \ac{md} spectrogram. The light grey points represent the raw measurements (observations) obtained as explained in \secref{sec:people-loc}, using Cartesian coordinates. The trajectory is correctly reconstructed with remarkable accuracy. The \ac{md} signature is extracted successfully and shows the different contributions of the torso and the limbs. The former reflects more power and follows a slightly oscillating motion, which is coherent with the direction changes in the walking trajectory, while the latter are responsible for the higher velocity peaks.

\begin{figure}[t!]
    \centering
    \includegraphics[width=4cm]{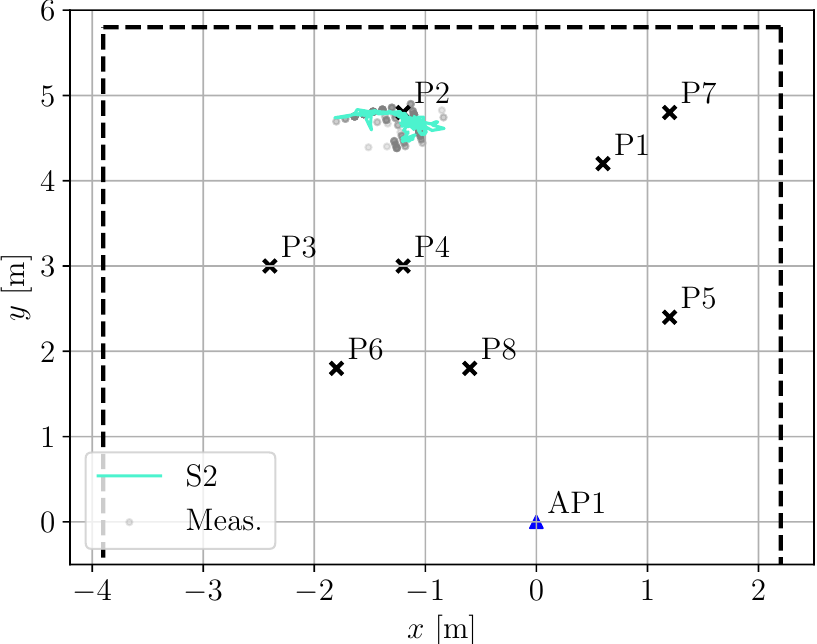}
    \includegraphics[width=4cm]{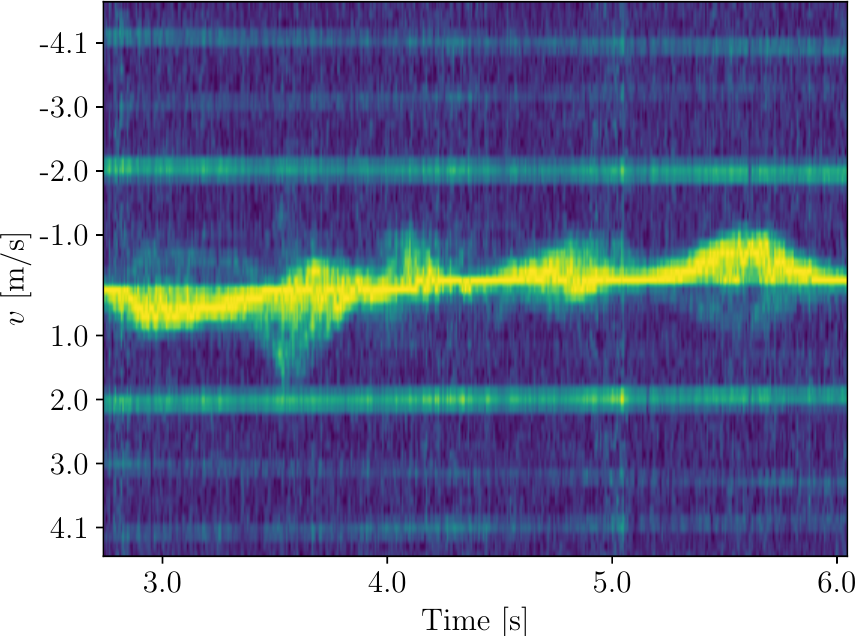}
    \caption{Estimated position of a subject sitting down (left) and a portion of the corresponding \ac{md} signature (right) extracted by RAPID.}
    \label{fig:sit_1sub}
\end{figure}

\begin{figure}[t!]
    \centering
    \includegraphics[width=7.5cm]{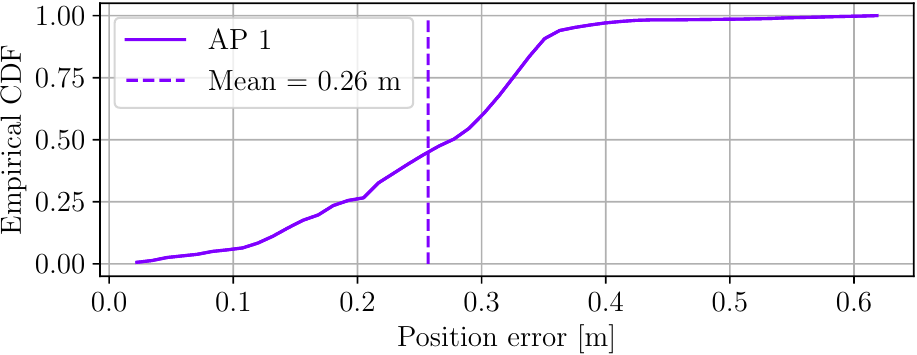}
    \caption{Empirical \ac{cdf} of the positioning error for a subject sitting down in correspondence of marker P$2$.}
    \label{fig:prel-cdf}
\end{figure}

Next, we test RAPID on a subject sitting down at the marker P$2$, as shown in \fig{fig:sit_1sub}. Also in this case, RAPID correctly estimates the location of the subject, and the \ac{md} spectrum is coherent with the sitting down activity. This is non-trivial, given that P$2$ is located at the edge of the experiment room. 
The empirical \ac{cdf} of the positioning error of the subject in \fig{fig:prel-cdf} shows that we achieve a good localization accuracy. In this analysis, we included around $2000$ position estimates made by the EKF. The median error is $26$~cm, and the probability of the error being lower than $40$~cm is close to $1$. We stress that the subject in this case is not static, as the person alternates between sitting down and standing up. This causes the estimated position to change slightly across time-steps, increasing the localization error.

Our baseline experiments empirically prove that IEEE~802.11ay Golay sequences are adequate for human tracking and \ac{md} extraction. This is not trivial, as: \textit{(i)} such sequences are not designed for sensing purposes and they have low Doppler resolution \cite{Kumari_11adradar_VTC}; \textit{(ii)} humans are believed to be poor reflectors of \ac{mmwave} 
signals, while we showed that a background subtraction step followed by \ac {aoa} estimation can reliably identify their contribution to the \ac{cir}.
While it is well known that human sensing can be performed with \ac{mmwave} radars employing frequency modulated \textit{chirp} signals \cite{vandersmissen2018indoor, pegoraro2021multiperson, meng2020gait}, RAPID is the first system to do so with \ac{mmwave} communication waveforms.


\subsection{Multi-person multi-AP tracking scenario}
\label{sec:multiperson-multiap}

In this section, we extend the scenario to analyze the impact of multiple subjects present on the scene, which we tackle using multiple \ac{ap}s.
Here, all measurements are performed using \ac{ap}$1$ and \ac{ap}$2$ in E$1$. We first consider the results obtained solely by \ac{ap}$1$, and then we combine \ac{ap}$1$ and \ac{ap}$2$.
Several experiments are carried out with $2$ to $5$ subjects, performing different activities.
In total, we collect $28$ such sequences each with duration $\sim 10$~s, of which $13$ include $2$ subjects, $5$ include $3$ subjects, $6$ include $4$ subjects and $4$ include $5$ subjects. \rev{These measurements are collected across different days, spanning a total of 3 weeks.}

\begin{figure}[t!]
    \centering
    \includegraphics[width=7cm]{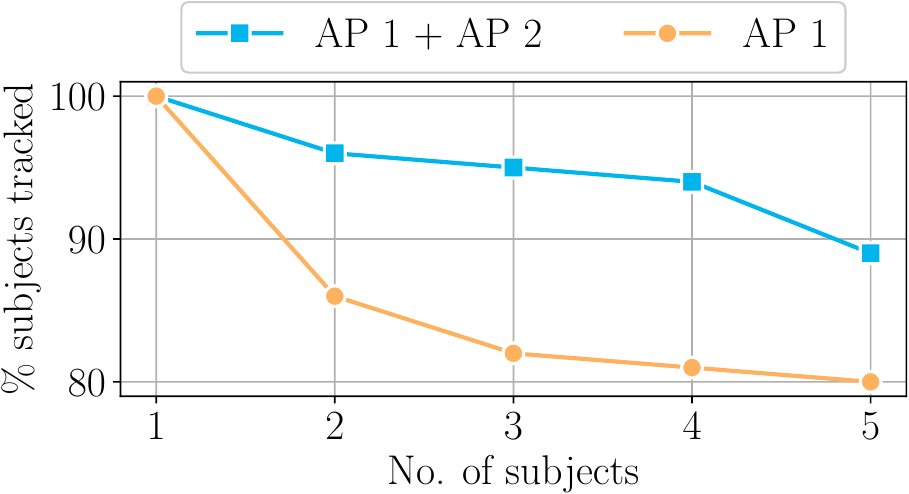}
    \caption{Rate of detection with a varying number of subjects using only \ac{ap}$1$ and the combination of \ac{ap}$1$ and \ac{ap}$2$.}
    \label{fig:det-ratio}
\end{figure}

\begin{figure*}[t!]
	\begin{center}   
		\centering
		\subcaptionbox{\label{fig:track_2sub}}[4cm]{\includegraphics[width=4cm]{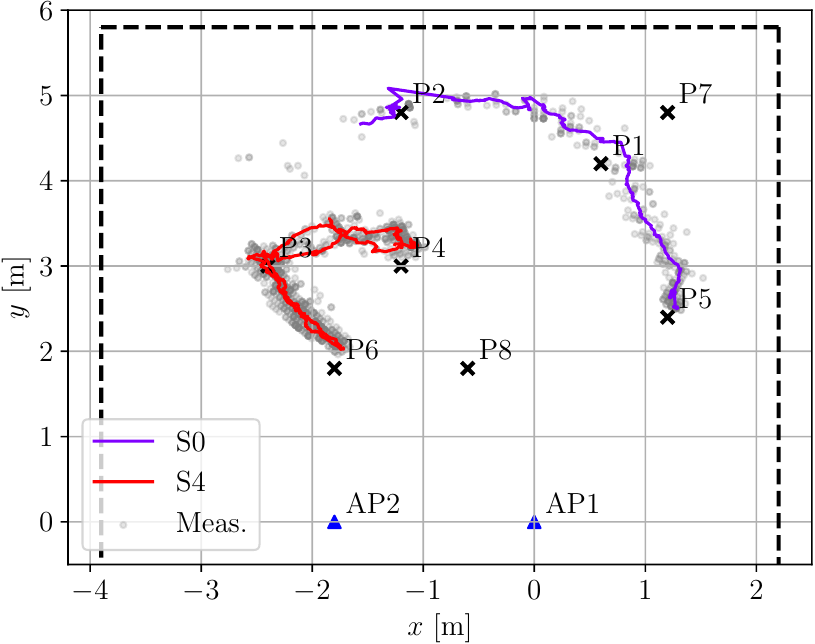}}
		\subcaptionbox{\label{fig:track_3sub}}[4cm]{\includegraphics[width=4cm]{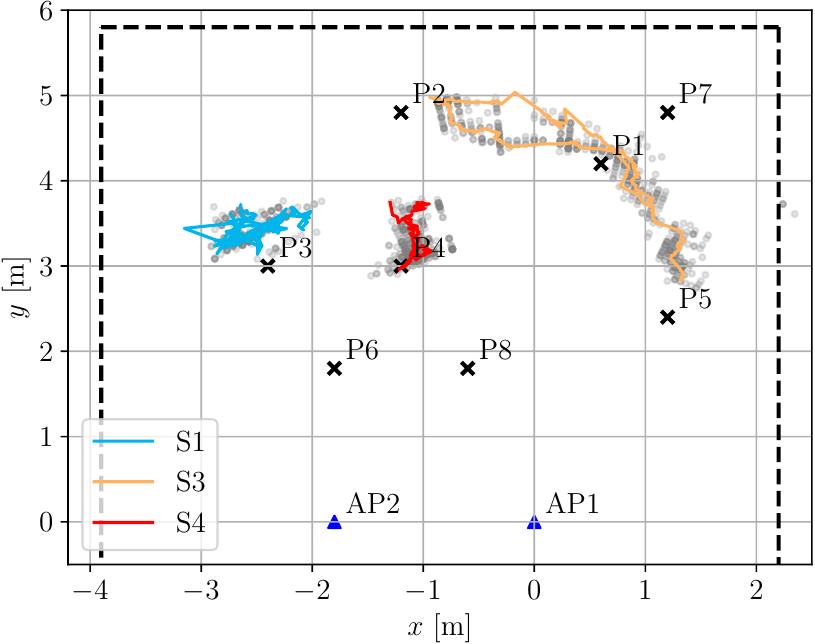}}
		\subcaptionbox{\label{fig:track_4sub}}[4cm]{\includegraphics[width=4cm]{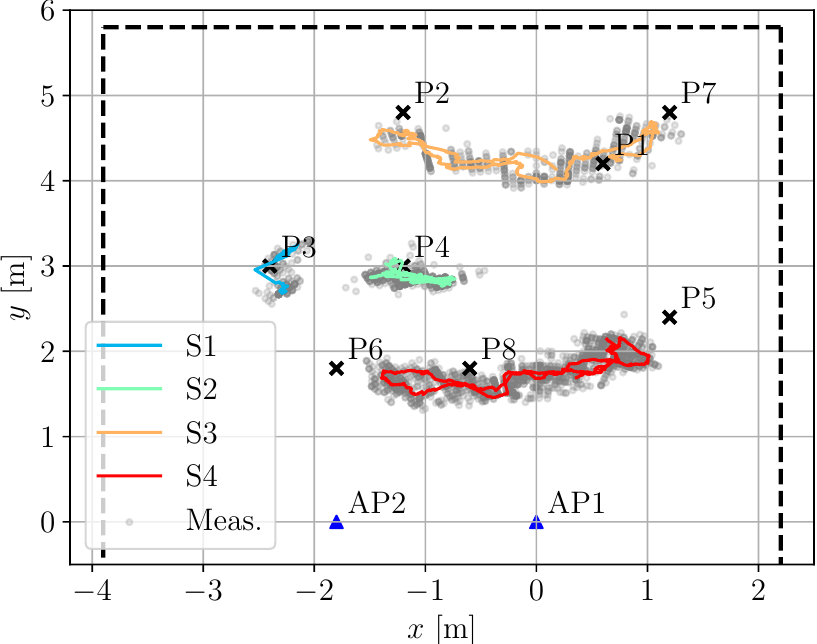}}
		\subcaptionbox{\label{fig:track_5sub}}[4cm]{\includegraphics[width=4cm]{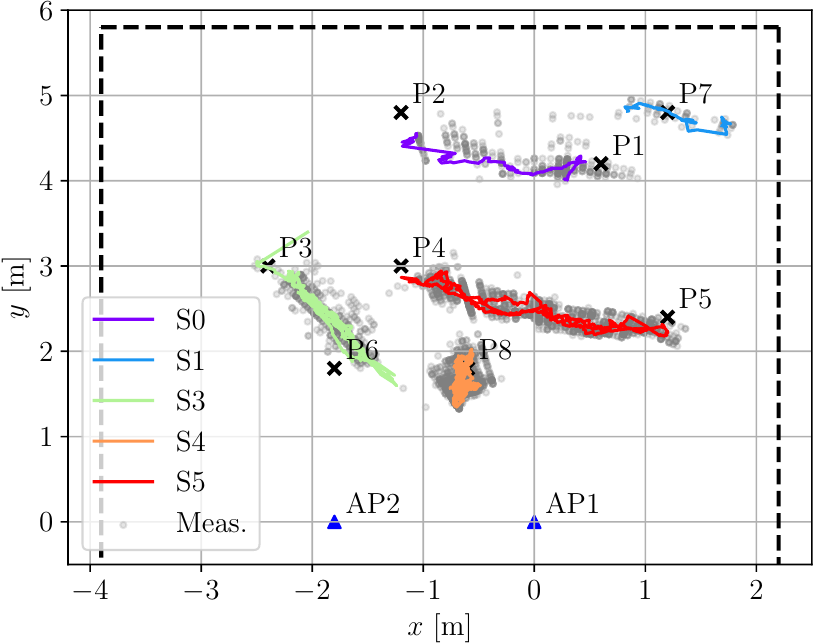}}
		\caption{EKF trajectories obtained in the multiperson scenario. Here a single \ac{ap} is used (\ac{ap}$1$). We show four successful cases in which RAPID is able to reconstruct the movement trajectories of $2$ \textbf{(a)}, $3$ \textbf{(b)}, $4$ \textbf{(c)} and $5$ \textbf{(d)} people moving the the room.}
		\label{fig:track-multi}
	\end{center}
 \vspace{-0.5cm}
\end{figure*}

\smallsection{Presence of multiple subjects} 
\fig{fig:track-multi} shows some example trajectories estimated by the EKF using the measurements from \ac{ap}$1$. 
RAPID is able to successfully track the users with considerable accuracy in most cases, even for $5$ subjects (see \fig{fig:track_5sub}). Note that this setup is extremely challenging, especially when more than $3$ subjects are present, due to the small dimensions of the environment that lead to a high probability of occlusion happening, i.e., one subject covers the \ac{los} path between the \ac{ap} and another individual.
\ac{mmwave} signals do not propagate through the human body, and occlusions may cause missed detection and tracking errors. On the other hand, in real-life scenarios occlusions may happen frequently, and the system must be robust to these events. In \fig{fig:det-ratio}, we report a quantitative analysis of the effect of increasing the number of subjects in terms of the percentage of subjects that are correctly detected and tracked by RAPID. Using only \ac{ap}$1$ we observe that, despite achieving adequate tracking performance, the system capability of detecting the subjects decreases significantly as their number increases. In particular, on average one subject goes undetected when $5$ individuals are present.

\begin{figure}[t!]
	\begin{center}   
		\centering
		\subcaptionbox{\ac{ap}$1$ estimated trajectories. \label{fig:occl-ap1}}[4cm]{\includegraphics[width=4cm]{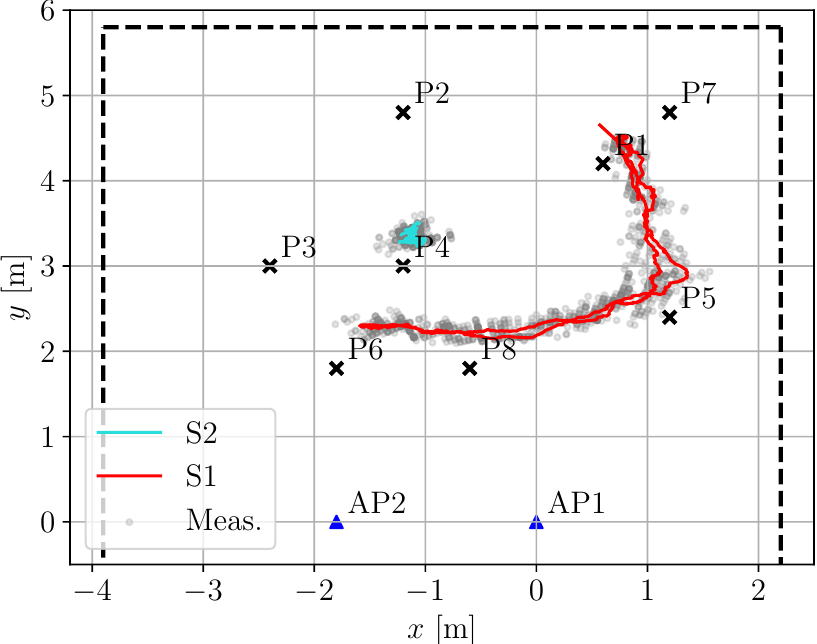}}
		\subcaptionbox{\ac{ap}$2$ estimated trajectories. \label{fig:occl-ap2}}[4cm]{\includegraphics[width=4cm]{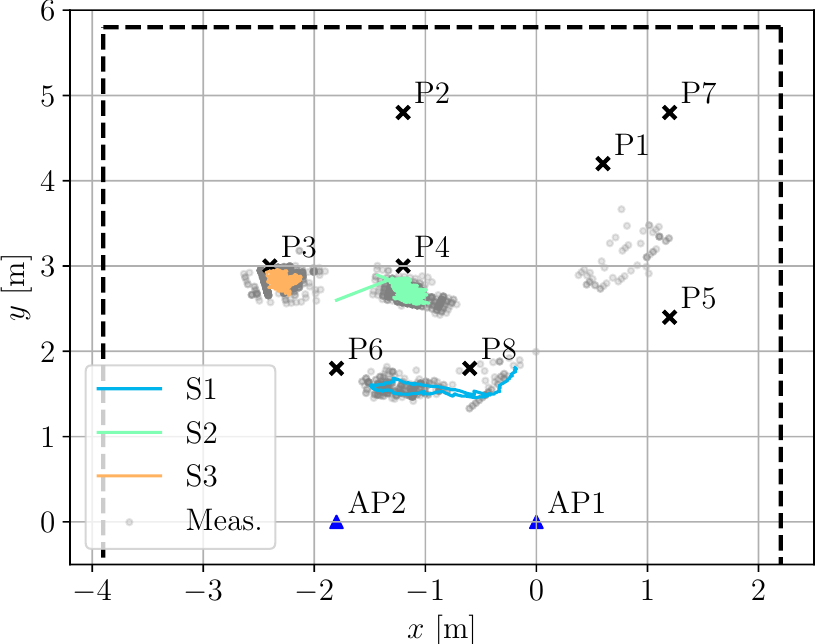}}
		\caption{Impact of using multiple \ac{ap}s on the occlusion problem. Here, \ac{ap}$1$ fails to detect and track S$3$, while \ac{ap}$2$ can only partially reconstruct the trajectory of S$1$. The combination of the $2$ \ac{ap}s successfully detects and tracks all subjects.}
		\label{fig:occlusion}
	\end{center}
\end{figure}
\begin{figure}[t!]
	\begin{center}   
		\centering
		\subcaptionbox{S$1$ running. \label{fig:running-p16}}[2.8cm]{\includegraphics[width=2.8cm]{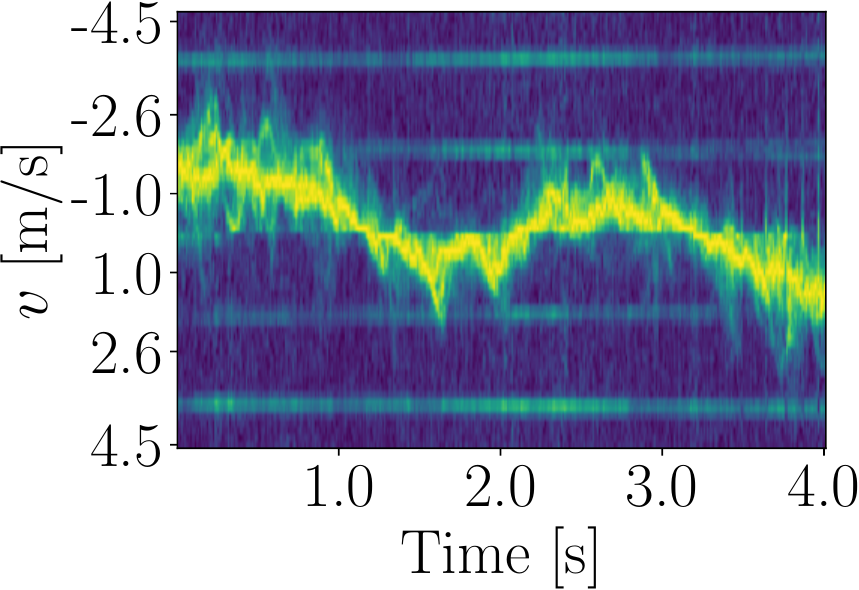}}
		\subcaptionbox{S$2$ sitting down. \label{fig:sitting-p4}}[2.8cm]{\includegraphics[width=2.8cm]{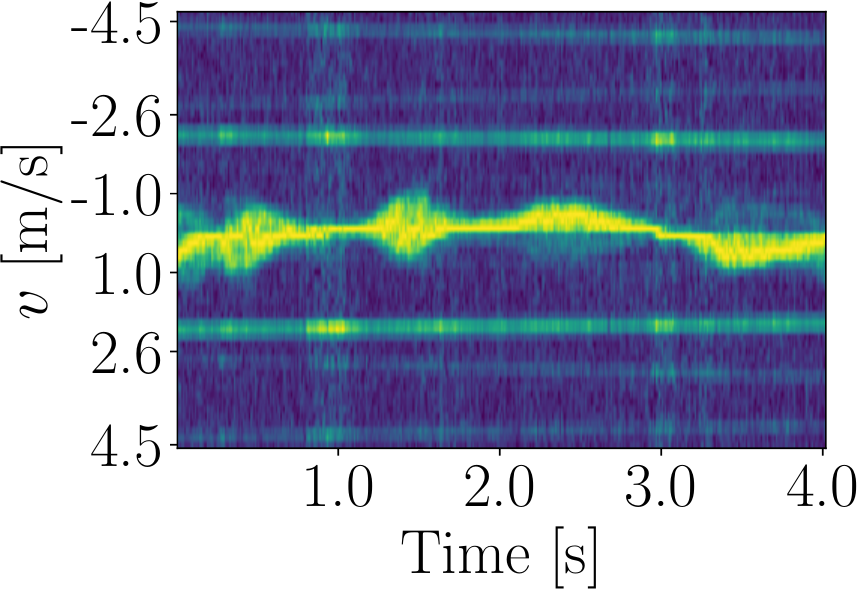}}
		\subcaptionbox{S$3$ waving. \label{fig:waving-p3}}[2.8cm]{\includegraphics[width=2.8cm]{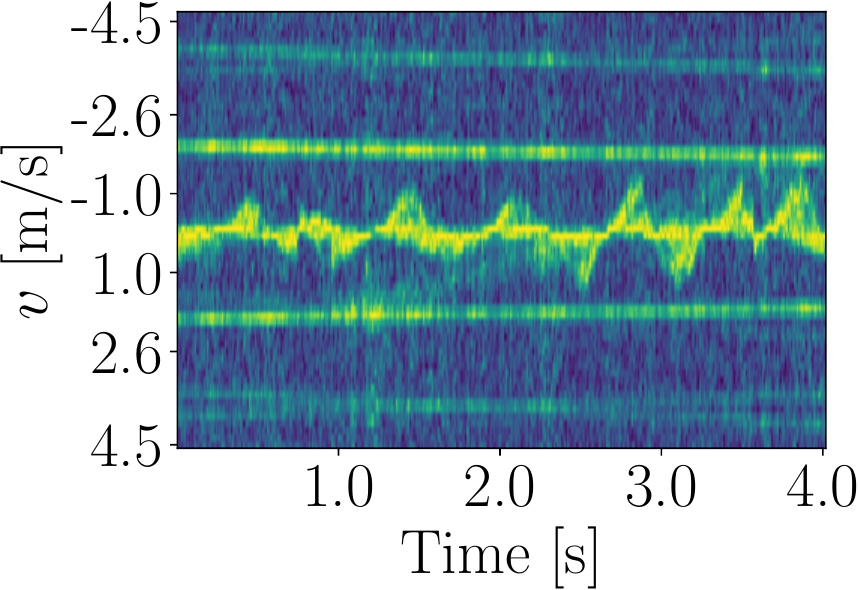}}
		\caption{Extracted \ac{md} signatures of the subjects in \fig{fig:occlusion}.}
		\label{fig:occlusion-md}
	\end{center}
 \vspace{-0.4cm}
\end{figure}
\begin{figure*}[t!]
	\begin{center}   
		\centering
         \includegraphics[width=8cm]{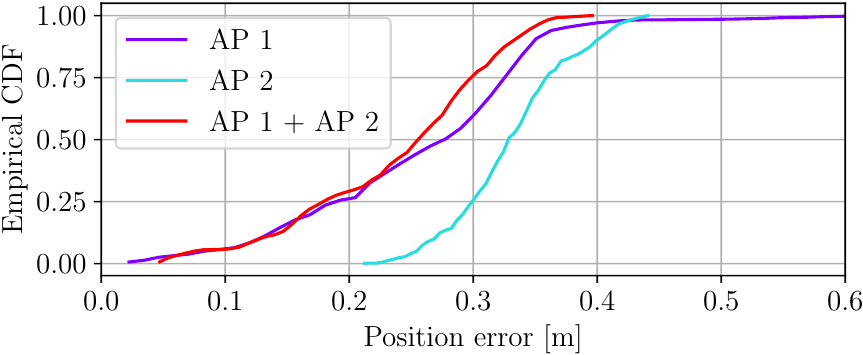}
		\includegraphics[width=8cm]{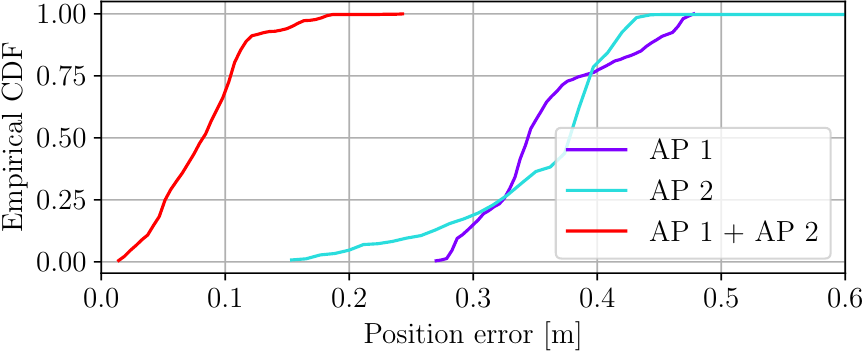}
		\caption{Localization error \ac{cdf}s for a subject sitting dwon in P$2$ (left) and in P$4$ (right). Combining multiple \ac{ap}s brings the largest improvement when their point-of-view on the subject is the most diverse.}
		\label{fig:loc-res}
	\end{center}
 \vspace{-0.5cm}
\end{figure*}
\smallsection{Improvement with multiple \ac{ap}s}
Combining the \acp{fov} of \ac{ap}$1$ and \ac{ap}$2$ effectively decreases the probability of occlusion events happening, as when the \ac{los} between an \ac{ap} and a subject is blocked, the other \ac{ap} can exploit its own \ac{los} path to detect the person. 
In \fig{fig:occlusion} we report a qualitative example of this, showing that RAPID  can effectively deal with occlusions by combining the \acp{fov} of the $2$ \ac{ap}s (in this case, $3$ subjects are present in the environment).
The EKF estimated trajectories from \ac{ap}$1$ are shown in \fig{fig:occl-ap1}: subjects S$1$ and S$2$ are successfully detected and tracked, while S$3$, who is waving hands in P$3$, is not. This is due to a combination of the occlusion caused by S$1$ and the fact that P$3$ is placed at the edge of the \ac{fov} of \ac{ap}$1$.
However, the position of \ac{ap}$2$ enables it to detect S$3$ successfully, while the trajectory of S$1$ can only be partially reconstructed.
Considering the trajectories estimated by both \ac{ap}s, RAPID can detect and track all subjects, successfully extracting their \ac{md} signatures, which are reported in \fig{fig:occlusion-md}. 

The subject detection rate is also significantly improved by using multiple \ac{ap}s, as shown in the blue curve in \fig{fig:det-ratio}. Despite \ac{ap}$1$ and \ac{ap}$2$ being placed along the same axis ($x$), and only $1.8$~m apart, this is sufficient to increase subject detection probability by $11\%$, $16\%$, $16\%$ and $11\%$ for the cases of $2$, $3$, $4$ and $5$ subjects, respectively.

Finally, we show the impact of \textit{averaging} the positions estimated by the two different \ac{ap}s, see \fig{fig:loc-res}. We repeat the experiment described in \secref{sec:prel-exp} with a single subject sitting down in position P$2$. Even using this simple fusion method, RAPID achieves a significant gain in the tail of the localization error distribution. A subject positioned in P$2$ represents a worst-case for this kind of analysis in our setting, as the locations of the \ac{ap}s with respect to this point are very similar in terms of distance and angle.
The same experiment is repeated for position P$4$, showing a larger improvement from combining the \ac{ap}s. In this case, RAPID goes from an average localization error of $0.35$~m using the single \ac{ap}s independently, down to an error of $0.08$~m by averaging their estimates. This is due to the more favorable positions from which P$4$ is illuminated by the \ac{ap}s.

\subsection{Impact of furniture and detection parameters}\label{sec:det-fa-results}

\begin{figure}[t!]
	\begin{center}   
		\centering
		\subcaptionbox{TR w/o. \label{fig:det-free}}[2.8cm]{\includegraphics[width=2.8cm]{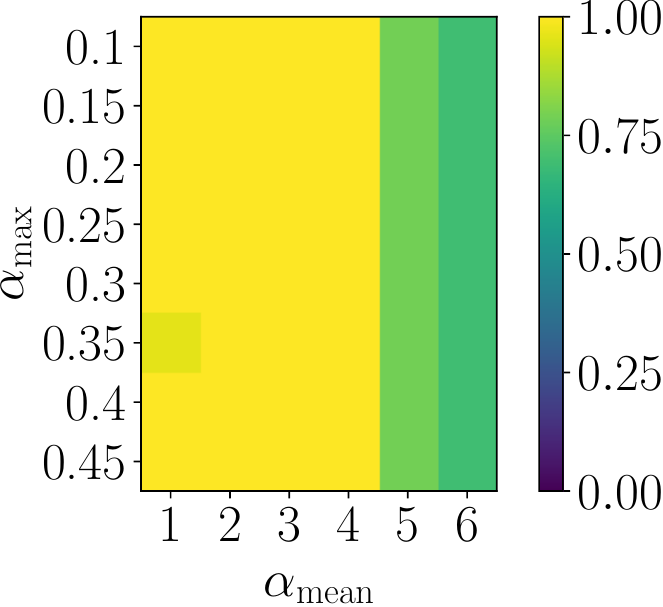}}
		\subcaptionbox{TR w/. \label{fig:det-table}}[2.8cm]{\includegraphics[width=2.8cm]{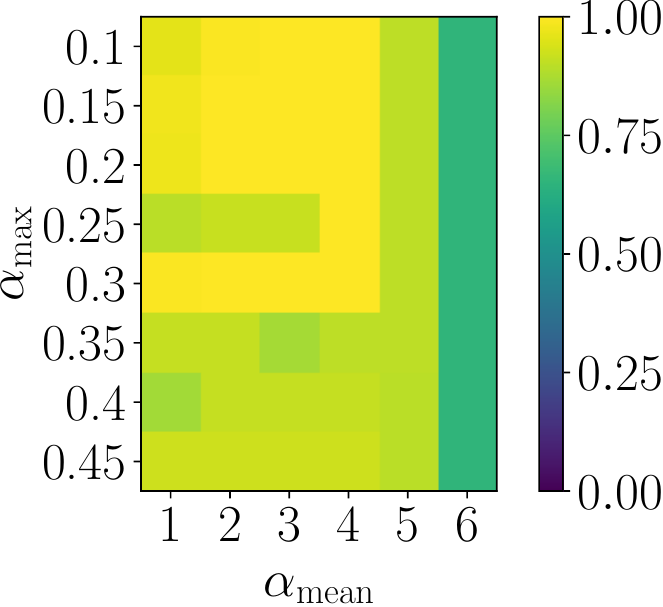}}
      \subcaptionbox{FR w/. \label{fig:ft-table}}[2.8cm]{\includegraphics[width=2.8cm]{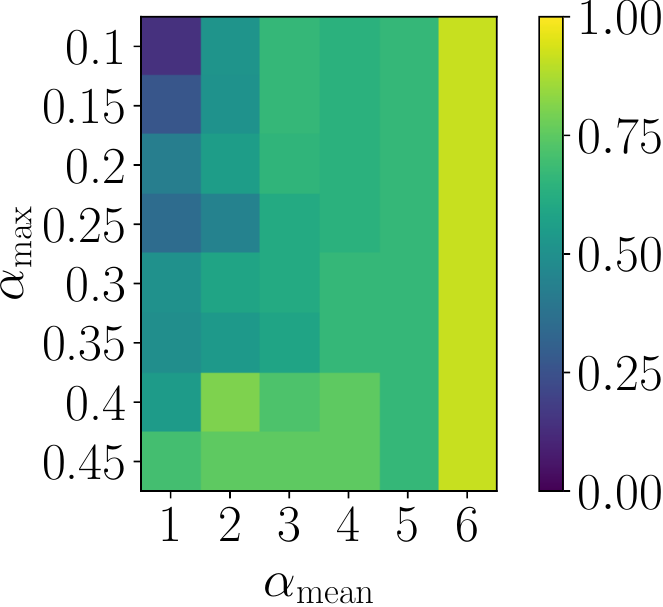}}
		\caption{Mean tracking rate (TR) and false track rejection rate (FR) obtained in a setup without furniture (w/o) and with a table (w/) placed in the sensing area. The heatmap shows the performance for different values of the parameters $\alpha_{\max}$ and $\alpha_{\rm mean}$.}
		\label{fig:det-params}
	\end{center}
\end{figure}

\begin{figure}[t!]
	\begin{center}   
		\centering
		\subcaptionbox{ \label{fig:pic-table}}[4.6cm]{\includegraphics[width=4.6cm]{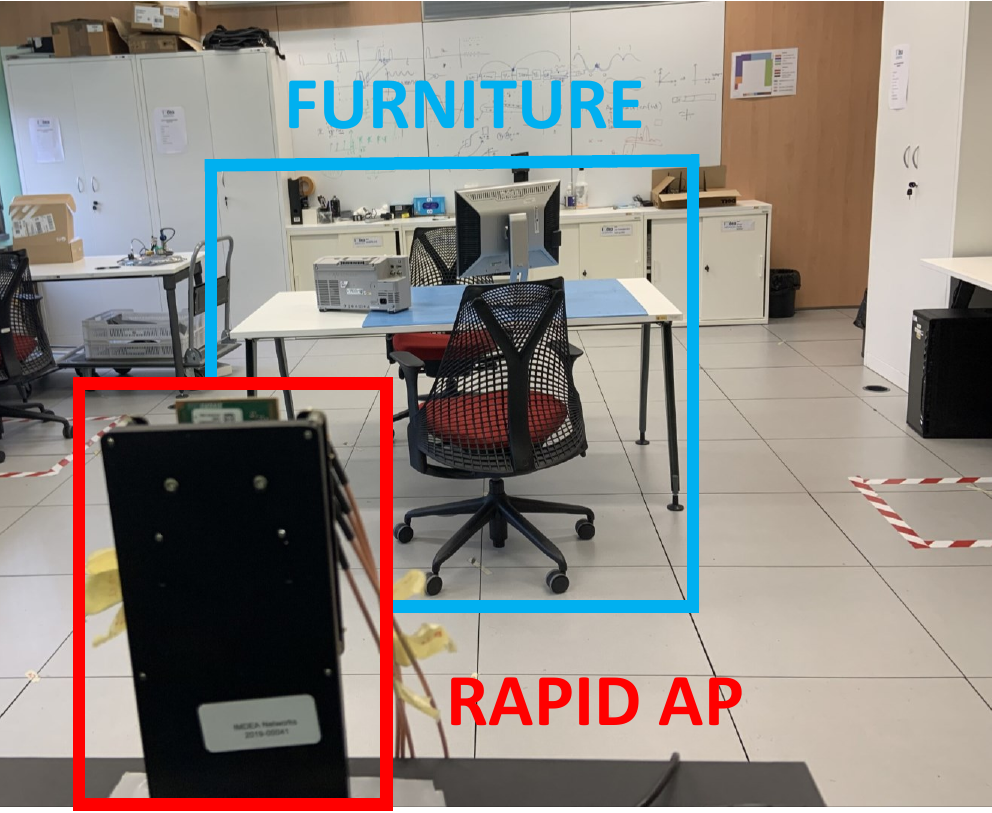}}
		\subcaptionbox{ \label{fig:map-table}}[3.9cm]{\includegraphics[width=3.9cm]{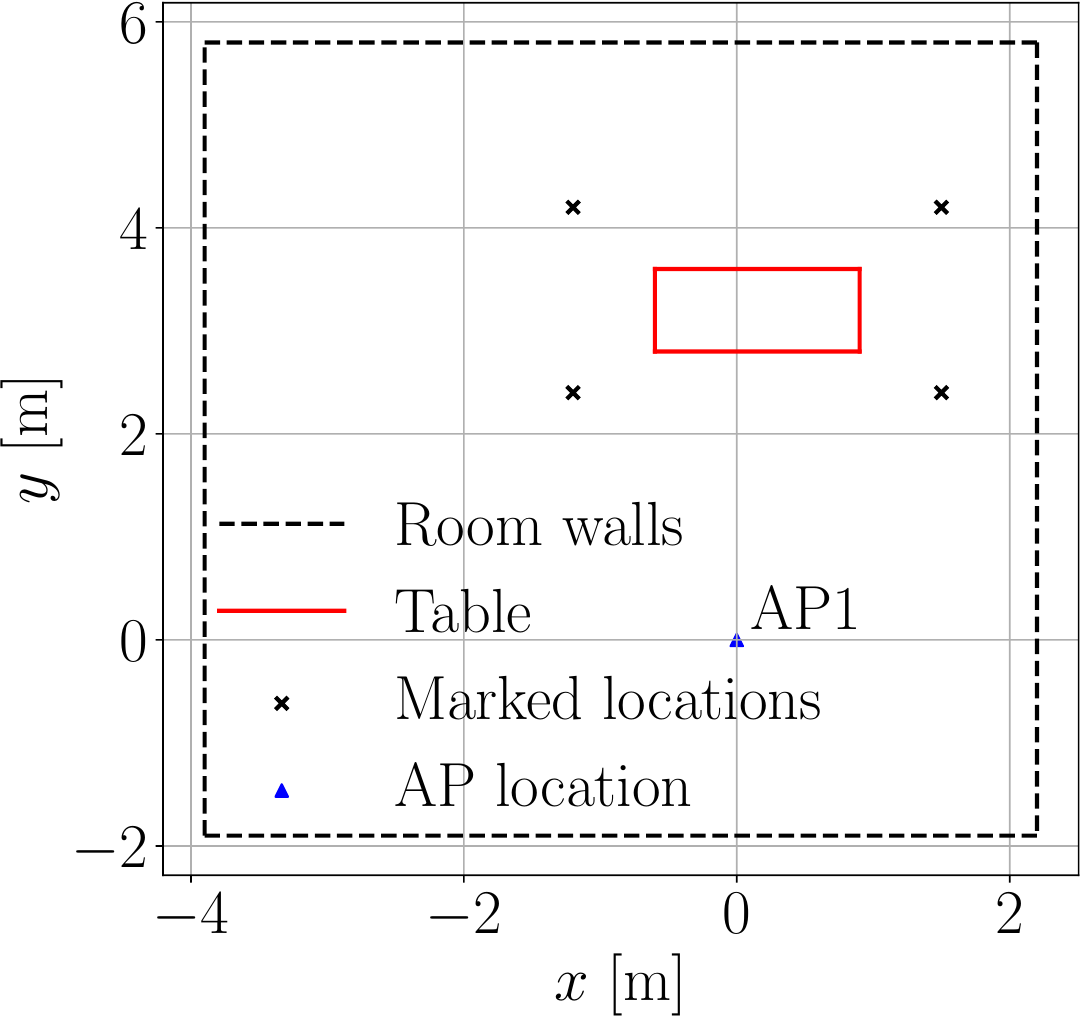}}
		\caption{E1 with furniture (a) and a schematic representation of the table location (b).}
		\label{fig:table}
	\end{center}
 \vspace{-0.4cm}
\end{figure}

In this section we analyze the impact of varying the main parameters of the proposed peak detection algorithm, $\alpha_{\max}$ and $\alpha_{\rm mean}$. To do so, we introduce the following two metrics: the tracking rate (TR) and the false tracks rejection rate~(FR). TR is defined as the fraction of time during which RAPID correctly tracks the subject. We consider a subject to be correctly tracked if the \ac{ekf} outputs a track that has an average tracking error with respect to the reference trajectory lower than $0.4$~m. FR is defined as the ratio $1/(N_{f} + 1)$, where $N_f$ is the number of spurious tracks outputted by the \ac{ekf}, i.e., those tracks not corresponding to the desired subjects. These can be generated due to false detections and/or reflections on background objects and furniture.
In \fig{fig:table}, we report the average TR and FR obtained by varying the parameters of the detection algorithm $\alpha_{\max}$ and $\alpha_{\rm mean}$ from $0.1$ to $0.45$ and from $1$ to $6$, respectively. \rev{The average is computed over $12$ measurement sequences, acquired on two different days, with a subject walking in the room along different trajectories.}
To evaluate the impact of furniture and obstacles between the RAPID \ac{ap} and the subject, we placed a table with a monitor, electronic equipment, and two chairs in the measurement space of E1, as shown in \fig{fig:table}. The subject was instructed to walk around and \textit{behind} the table across the four markers shown in \fig{fig:map-table}.
\fig{fig:det-free} and \fig{fig:det-table} contain the TR without (w/o) and with furniture (w/). Notice how the presence of obstacles reduces the range of parameters that lead to good TR. 
To select adequate $\alpha_{\max}$ and $\alpha_{\rm mean}$, one has to strike a balance between a high TR, which ensures the target is reliably detected and tracked, and high FR, which indicates that the number of false tracks created is low.
Lowering the detection parameters yields high TR, but leads to the creation of more undesired tracks, as the sensitivity of the detection is increased. This is shown in \fig{fig:ft-table}, where we plot the average FR varying the detection parameters in a setup with furniture. Combining the three heatmaps in \fig{fig:det-params}, one can see that suitable values of $\alpha_{\max}$ are between $0.1$ and $0.3$, while for $\alpha_{\rm mean}$ we suggest $3$ or $4$ to avoid generating too many spurious tracks. 
In the following results, we used $\alpha_{\max}=0.15$, $\alpha_{\rm mean}=3$.

\rev{Next, we compute the absolute tracking error between the \ac{ekf} output trajectory and the ground truth path passing through the four markers. These values are reported in \tab{tab:track-rmse} in the case of no furniture in the room (w/o) and with furniture (w/), along with the corresponding standard deviations. Occlusions due to furniture only slightly degrade the tracking accuracy ($5$~cm higher error). This is is due to the fact that, even when furniture is present, RAPID can at least detect the main reflection from the subject's torso most of the time, obtaining precise estimates for the distance between the \ac{ap} and the person. These are then smoothed across time by the \ac{ekf}, yielding an accurate trajectory. The last column in \tab{tab:track-rmse} reports the tracking error limited to the part of the estimated trajectory where the subject is completely occluded, so the \ac{ekf} outputs linear predictions based on past measurements until the subject becomes detectable again. This leads to a noticeable (but still contained) degradation of the tracking performance, which is however expected to drop even further in case of more complex, non-linear movement trajectories.}

\begin{table}[t!] 
	\caption{\rev{Tracking RMSE without (w/o) and with (w/) furniture, and under complete occlusion of the subject.}} \label{tab:track-rmse}
    \vspace{-0.25cm}
	\begin{center}
		\begin{tabular}{lccc}
			\toprule	
			&w/o furniture &w/ furniture & compl. occlusion\\
			\cmidrule(lr){2-4}
			\textbf{RMSE [cm]} &$20.0 \pm 6.3 $&$25.1 \pm 8.0$& $40.9 \pm 11.9$\\
			\bottomrule
		\end{tabular} 		
	\end{center}
\end{table}

\subsection{Human activity recognition}\label{sec:act-rec-res}
\begin{figure}[t!]
    \centering
    \includegraphics[width=3.8cm]{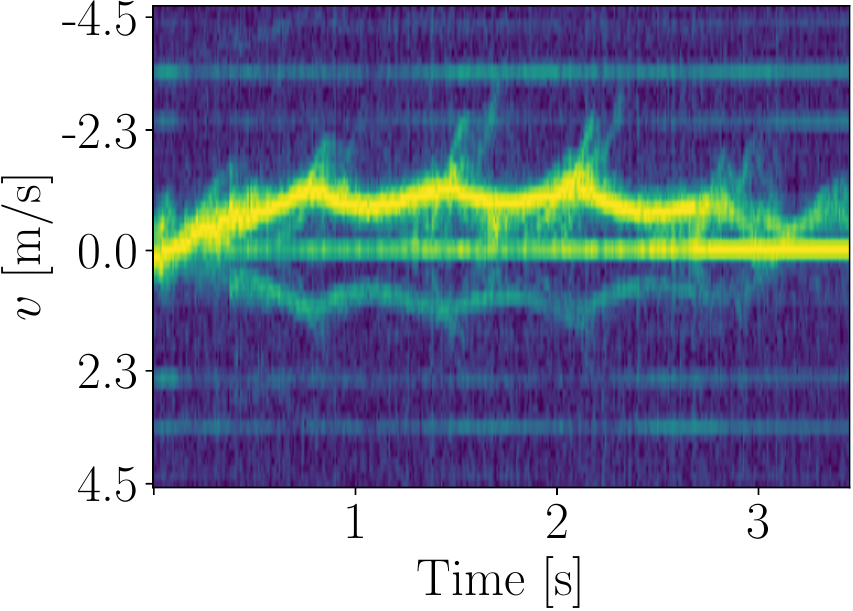}
    \includegraphics[width=3.8cm]{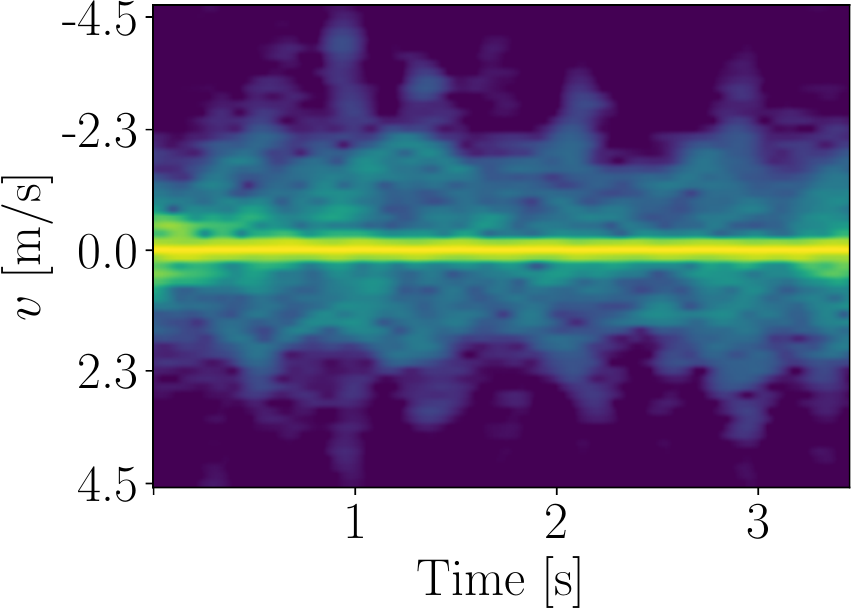}
    \caption{Walking spectrogram concurrently obtained with RAPID at $60$~GHz (left) and with sub-$6$~GHz sensing (right).}
    \label{fig:mmwave-sub6-spec}
\end{figure}
Next, we evaluate the \ac{har} performance of RAPID, comparing it to legacy sub-$6$~GHz WiFi systems.
For all the experiments in this section, unless stated otherwise, we used a unique labeled training dataset of simultaneous IEEE~802.11ay \ac{cir} (at $60$~GHz) and IEEE~802.11ac \ac{cfr} (at $5$~GHz) sequences, which we collected in E$1$, with a single subject performing the $5$ different activities A$0-4$. We used a single RAPID \ac{ap} and a pair of transmitter/receiver IEEE~802.11ac routers with $4$ antenna elements (ASUS RT-AC$86$U implementing the Nexmon-CSI firmware modifications~\cite{Gringoli2019}). 
The estimates are obtained with the two systems operating \textit{(i)} \textit{concurrently}, i.e., each training/testing sequence for the same activity of the subject is collected with both the RAPID \ac{mmwave} \ac{ap} and the sub-$6$~GHz system, and \textit{(ii)} with the same \ac{md} frequency range and resolution. The latter is achieved by tuning the IEEE~802.11ac system inter-packet transmission time using a slight modification of  \eq{eq:dvel-res-range} for the case of non co-located transmitter and receiver, i.e., $\Delta v = c / (f_o^{\rm ac} M T_c^{\rm ac})$ with $f_o^{\rm ac} = 5$~GHz. Therefore, the IEEE~802.11ac inter-packet transmission time is computed as $T_c^{\rm ac} = 2 T_c f_o / f_o^{\rm ac} \approx 6$~ms. The data are obtained in sequences of approximately $10$~s, for a total of around $6$~minutes of \ac{cir}/\ac{cfr} measurements per activity. \rev{Those sequences are gathered on multiple days over the course of one month.}
Next, the \ac{md} spectrograms are obtained from the collected data.
To do this in the sub-$6$~GHz system, we adopt the pre-processing steps proposed in \cite{Zheng2019}, to which we refer for additional details.

The resulting \ac{md} spectrograms are split into partially overlapping windows of $1.728$~s, which are the input to the \ac{cnn}. For RAPID, we use windows containing $N_{\mu{\rm D}}=200$ time-steps while for the sub-$6$~GHz setup each window consists of $287$ samples. In \fig{fig:mmwave-sub6-spec} we show an example of the \ac{md} signatures obtained by RAPID and by the sub-$6$~GHz system for the same measurement sequence of a walking person.
We use the \ac{cnn} model detailed in Section~\ref{sec:dl-classifier} for both mmWave and sub-$6$~GHz spectrograms.
The \ac{cnn} is trained using the cross-entropy loss function \cite{goodfellow2016deep} and the Adam optimizer \cite{kingma2015adam}, with learning rate $10^{-4}$, until convergence of the loss function on a subset of the training data, used as validation set. 
We evaluate the performance of the classifier with a weighted average of the per-class F$1$-score metric, based on the number of samples per class. The F$1$-score is defined as $\texttt{tp} / [\texttt{tp} + 0.5(\texttt{fp} + \texttt{fn})]$, where \texttt{tp}, \texttt{fp} and \texttt{fn} are the predicted true positives, false positives and false negatives, respectively.

\noindent \smallsection{Single person, single AP scenario}
In \tab{tab:har-single-rapid} we report the confusion matrix and per-class F$1$-scores obtained by RAPID (grey rows) and by the IEEE~802.11ac system (white rows) on test sequences containing data from the same subject present in the training set, collected in E$1$. This evaluation is also referred to as our \textit{baseline} \ac{har} experiment in the following. Comparing the two systems, one can see that RAPID accurately classifies all activities, only showing slightly lower performance on A$2$, sitting down, as this mostly involves body movements directed along an orthogonal direction with respect to the receiver (along the vertical axis). Indeed, the motion-induced \ac{md} phase displacement is only measurable in the \textit{radial} direction as we rely on the direct path between the subject and the \ac{ap}. Sub-$6$~GHz, instead, benefits from a richer multipath environment and better recognizes A$2$, but confuses the other activities, especially walking with running and standing still. This is due, in part, to the low resolution of the \ac{md} obtained at $5$~GHz, which contains coarser-grained information (see  \fig{fig:mmwave-sub6-spec}).

\begin{table}[t!] 
	\caption{Confusion matrix and F$1$-scores for the baseline case. Grey/white rows refer to RAPID and sub-$6$~GHz, respectively.} \label{tab:har-single-rapid}
    \vspace{-0.25cm}
	\begin{center}
		\begin{tabular}{lccccc}
			\toprule	
			&\multicolumn{5}{c}{{\bf Predictions $[\%]$}}\\
			\cmidrule(lr){2-6}
            \vspace{0.15cm}
			\textbf{True $[\%]$}&Walking&Running&S. down&Waving&Still\\
			\multirow{ 2}{*}{Walking} & $\boldsymbol{97.7}$ \cellcolor[HTML]{ebebeb}& $0$\cellcolor[HTML]{ebebeb} & $2.3$ \cellcolor[HTML]{ebebeb}& $0$ \cellcolor[HTML]{ebebeb}& $0$\cellcolor[HTML]{ebebeb}\\
			\vspace{0.15cm}
			                           & $61.4$ & $18.3$ & $0$ & $0$ & $20.3$\\
			\multirow{ 2}{*}{Running} &$0$\cellcolor[HTML]{ebebeb}&$\boldsymbol{100}$\cellcolor[HTML]{ebebeb}&$0$\cellcolor[HTML]{ebebeb}&$0$\cellcolor[HTML]{ebebeb}&$0$\cellcolor[HTML]{ebebeb}\\
			\vspace{0.15cm}
		                              &$0.6$&$87.1$&$0$&$0$&$12.3$\\
			\multirow{ 2}{*}{S. down} &$0$\cellcolor[HTML]{ebebeb}&$0$\cellcolor[HTML]{ebebeb}&$95.9$\cellcolor[HTML]{ebebeb}&$0$\cellcolor[HTML]{ebebeb}&$4.1$\cellcolor[HTML]{ebebeb}\\
			\vspace{0.15cm}
    	                              &$0$&$0$&$\boldsymbol{100}$&$0$&$0$\\
			\multirow{ 2}{*}{Waving} &$0$\cellcolor[HTML]{ebebeb}&$0$\cellcolor[HTML]{ebebeb}&$0$\cellcolor[HTML]{ebebeb}&$\boldsymbol{100}$\cellcolor[HTML]{ebebeb}&$0$\cellcolor[HTML]{ebebeb}\\
			\vspace{0.15cm}
		                             &$0$&$0$&$0.1$&$85.9$&$14.0$\\
			\multirow{ 2}{*}{Still} &$0$\cellcolor[HTML]{ebebeb}&$0$\cellcolor[HTML]{ebebeb}&$0$\cellcolor[HTML]{ebebeb}&$0$\cellcolor[HTML]{ebebeb}&$100$\cellcolor[HTML]{ebebeb}\\
	  	                            &$0$&$0$&$0$&$0$&$100$\\
	  	                            \midrule
	  	                    \multirow{ 2}{*}{\textbf{F1-score}  $[\%]$} &
	  	      $\boldsymbol{98.8}$ \cellcolor[HTML]{ebebeb}& $\boldsymbol{100}$ \cellcolor[HTML]{ebebeb}&  $96.3$ \cellcolor[HTML]{ebebeb}& $\boldsymbol{100}$\cellcolor[HTML]{ebebeb} &  $\boldsymbol{98.4}$\cellcolor[HTML]{ebebeb} \\              & $75.8$&$91.2$ & $\boldsymbol{99.9}$& $92.4$ & $85.4$\\ 
			\bottomrule
		\end{tabular} 		
	\end{center}
\end{table}

\begin{figure}
    \centering
    \includegraphics[width=0.9\linewidth]{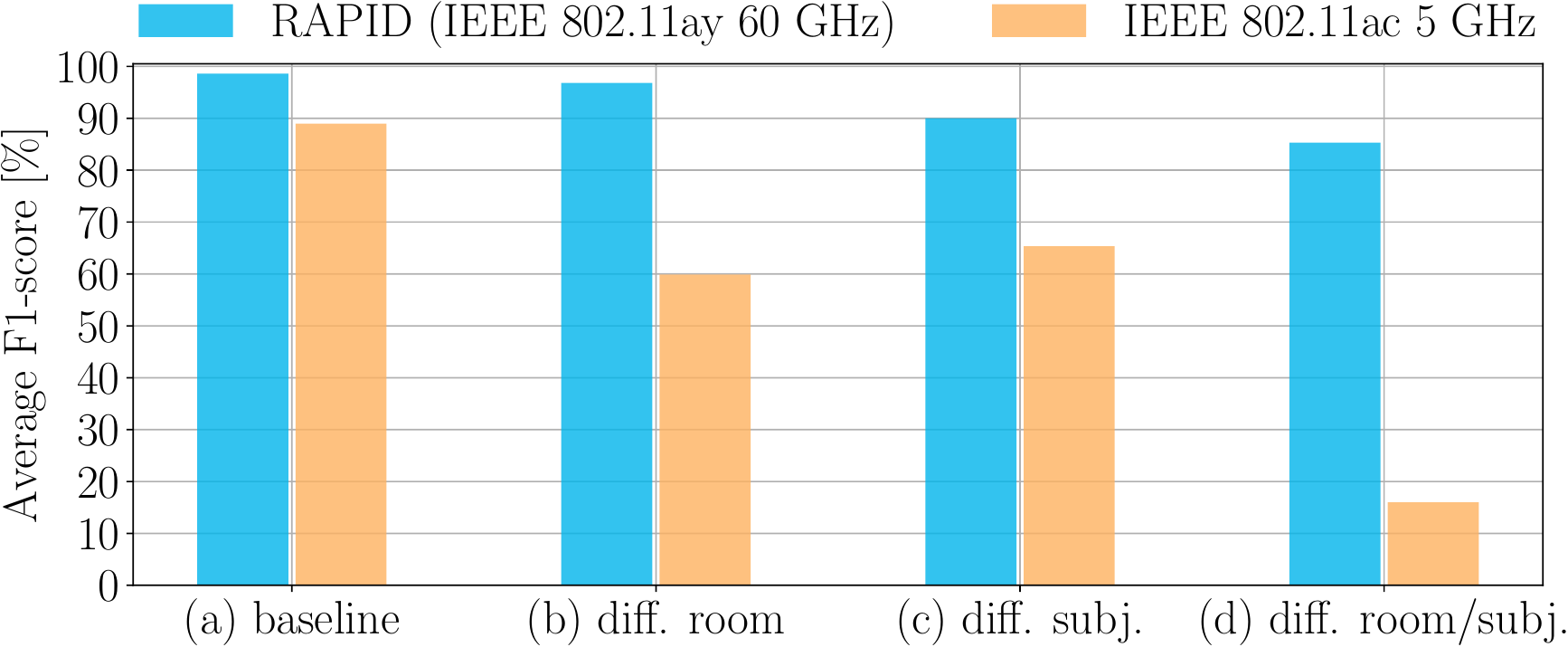}
    \caption{Comparison between the \ac{har} F$1$-score obtained by RAPID and by standard IEEE~802.11ac sensing at $5$~GHz for various scenarios.}
    \label{fig:mmwave-sub6}
\end{figure}

\noindent \smallsection{Impact of unknown environment and subject} Next, we further evaluate the \ac{har} robustness of the two systems in more complex settings, involving a different room than the one used for the training data collection (E$2$), and a different subject performing the activities. \fig{fig:mmwave-sub6} reports the weighted average of the per-class F$1$-scores obtained with RAPID and the sub-$6$~GHz system: (a) in the baseline scenario, (b) in a different room, E$2$, on the same subject (c) with a different subject, in the same environment (E$1$) and (d) in a different environment (E$2$) and on a different subject. The results show that RAPID outperforms the sub-$6$~GHz counterpart in generalizing to new environments and subjects, showing much lower performance degradation when moving to an unknown room or testing on a different person. 
In scenario (d) the sub-$6$~GHz \ac{har} system completely fails, obtaining a very low F$1$-score, due to the challenging combination of a different room and a different subject. Conversely, RAPID still achieves good performance. We stress that here the training data contain measurements from only one subject. Therefore, the \ac{cnn} classifier must possess great generalization capabilities to correctly classify the activities performed by another person, as they may have slightly different features.

In addition, we test the two systems under \textit{interference} from another subject in one of the activities of the training set, as shown in \tab{tab:interf-comp}. For this, we use the same setting as in the baseline, but we replace the training data for A$3$, waving hands, with new measurements where another person, termed \textit{interfering subject}, is present in the room besides the subject performing A$3$. The interfering subject performs a different, randomly selected, activity in each measurement sequence, in a position close to the intended subject, thus possibly disturbing the useful signal reflections. RAPID, thanks to the separation between different subjects enabled by the high ranging accuracy of mmWaves and the tracking process, is highly robust to the presence of other people. Sub-$6$~GHz sensing, instead, suffers from its low ranging resolution ($\sim 4$~m) and is greatly affected by the interference. 

\begin{table}[t!] 
	\caption{\ac{har} performance under interference from another subject in the training dataset.} \label{tab:interf-comp}
    \vspace{-0.25cm}
	\begin{center}
		\begin{tabular}{lccccc}
			\toprule	
			\textbf{F1-score $[\%]$}&Walking&Running&S. down&Waving&Still\\
			\cmidrule(lr){2-6}
			RAPID &$\boldsymbol{98.5}$& $\boldsymbol{99.9}$ &  $93.2$ &$\boldsymbol{100}$& $\boldsymbol{96.6}$\\              
			Sub-$6$~GHz & $72.2$ & $92.4$ & $\boldsymbol{97.8}$ & $58.0$ & $77.8$\\
			\bottomrule
		\end{tabular} 		
	\end{center}
\end{table}

\noindent\smallsection{Multi-person, multi-AP scenario} 
Next, we evaluate RAPID's \ac{har} performance degradation when multiple subjects are concurrently present in the environment, each performing, in general, a different activity. The aim here is to assess the effectiveness of RAPID in the separation of \ac{md} signatures associated with different targets.
In this evaluation, we do not consider the sub-$6$~GHz system, as the intrinsic limits in terms of ranging ($\sim 4$~m) and angular ($\sim 20^{\circ}$) resolutions prevent people tracking in crowded indoor scenarios such as the ones under study \cite{korany2020multiple}, thus making the separation of the multiple subjects infeasible.

We collect a labeled training dataset including $6$ subjects performing the $5$ different activities A$0-4$ using a single RAPID-\ac{ap}. The data are obtained in sequences of approximately $10$~s, and the resulting \ac{md} spectrograms are split into windows of $1.728$~s as in the single target case. \rev{In total, this dataset contains around $2$~minutes per activity \textit{per subject} split into multiple captures. These are acquired on different days, over the course of 3 weeks.} 

By training on different subjects, we aim at mitigating the \ac{har} performance reduction due to the difficulty of generalizing to different people, to better gauge the sole effect of \ac{md} separation.
We test the trained model on the same multi-person sequences used in \secref{sec:multiperson-multiap}, adding $6$ additional sequences with a single subject, for a total of $34$ sequences.
We use the RAPID processing steps to extract the \ac{md} signatures of each subject's movement; when using $2$ \ac{ap}s, we use the decision fusion scheme from \secref{sec:dec-fus}. 

\tab{tab:act-2ap-gain} shows the F$1$-score of RAPID for a varying number of people in the scene, and the gain obtained by combining the $2$ \ac{ap}s with respect to using only \ac{ap}$1$. In addition, we also report the corresponding detection rate, previously shown in \fig{fig:det-ratio}, for completeness. We observe that the F$1$-score only slightly decreases when moving from $2$ to $5$ subjects. This shows that the proposed \ac{md} extraction process can reliably separate the contributions of the different individuals.
In addition, combining multiple \ac{ap}s can bring a slight improvement in some cases, by exploiting the different illumination angles of the devices.

\begin{figure}[t!]
    \centering
    \includegraphics[width=0.8\linewidth]{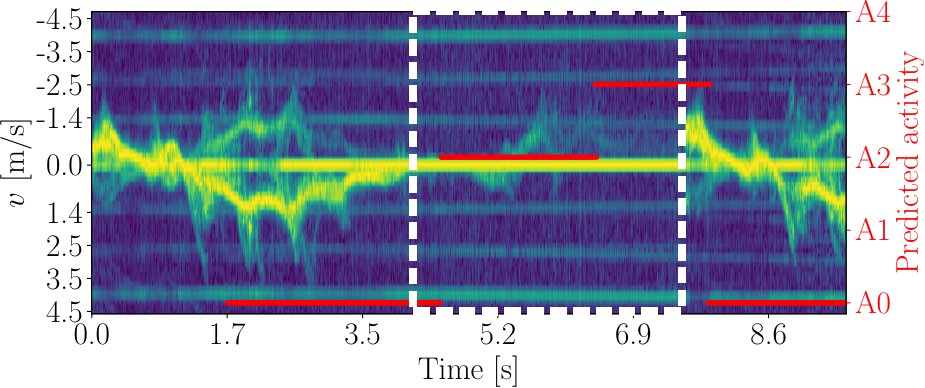}
    \caption{\rev{\ac{md} signature of a walking subject (A0) under temporary occlusion due to the presence of furniture. The dashed white rectangle highlights the occlusion period, while red lines indicate RAPID's predicted activity in each frame.}}
    \label{fig:eval-occlusion-har}
\end{figure}

\noindent\rev{\smallsection{Impact of occlusions due to furniture} Finally, we evaluate the impact of the presence of furniture on the \ac{har} task, using the same experimental setup described in \secref{sec:det-fa-results} (see  \fig{fig:table}). We showed previously that RAPID's human tracking and detection is only slightly affected by such occlusions, as the main \ac{cir} peaks are still detectable. However, \ac{har} is much more challenging as the quality of the \ac{md} signature may be significantly degraded even by \textit{partial} body occlusions, as the key contribution of some body parts (e.g., legs and/or arms) may not be visible in the spectrogram. In \fig{fig:eval-occlusion-har}, we show the \ac{md} obtained from a subject walking (A0) around the table in \fig{fig:table}. The additional $y$-axis on the right and the red lines represent the activity predicted by RAPID, obtained by sweeping the $N_{\rm \mu D}$ frames long \ac{cnn} input window over the \ac{md}. After an initialization time needed to collect the first window, RAPID correctly classifies activity A0. During the subsequent occlusion event (enclosed in the dashed white rectangle), the torso reflection in the \ac{md} becomes much fainter, while the contributions of the other body parts disappear, causing misclassifications (A2 and A3). We stress that this is due to an intrinsic limitation of \ac{mmwave} signals. In fact, in case of occlusion the information about the different body parts is mostly \textit{undetectable} at the receiver, hence the available information about the movement is insufficient to correctly classify it. However, RAPID promptly re-establishes the correct classification when the subject becomes visible again. Note that RAPID is not trained with samples including occlusion events in the training set, which makes this test even more challenging. 
}

\begin{table}[t!] 
	\caption{\ac{har} F$1$-score and detection rate  vs. no. of concurrent users. 
} \label{tab:act-2ap-gain}
	\begin{center}
		\begin{tabular}{ccccccc}
			\toprule	
			\textbf{\ac{ap}s}&\textbf{Metric}&$1$ subj.&$2$ subj.&$3$ subj.&$4$ subj.&$5$ subj.\\
			\cmidrule(lr){1-2}\cmidrule(lr){3-7}
			\multirow{ 2}{*}{$1$}&F$1$&$99.9$&$99.3$&$97.9$&$95.3$&$94.4$\\
			\vspace{0.15cm}
			&Det. rate&$100$&$86.1$&$82.9$&$81.3$&$80.0$\\
			\multirow{ 2}{*}{$1$ \& $2$}&F$1$&$100$&$99.4$&$99.4$&$95.4$&$94.4$\\
			&Det. rate&$100$&$96.7$&$95.5$&$94.5$&$89.2$\\
			\bottomrule
		\end{tabular} 		
	\end{center}
 \vspace{-0.4cm}
\end{table}

\subsection{Person identification}\label{sec:p-id-res}

 In this section we test the performance of RAPID on person identification, by building a dataset including the gait \ac{md} spectrograms of $7$ subjects, collected in E$1$. \rev{We collect from $3$ to $5$~minutes of training data per subject, split into multiple captures acquired over the course of 10 days. The data from each subject is the collection of captures obtained on different days, to avoid slight daily variations in the gait to bias the dataset.} The input samples for the classifier are obtained using \ac{md} windows of the same length as for \ac{har}, i.e., $1.728$~s. The \ac{cnn} classifier is trained using the same parameters and loss function used for \ac{har}.

\smallsection{Person identification accuracy}
First, we evaluate the accuracy of person identification on a varying number of subjects to recognize.
In \tab{tab:person-id} we report the accuracy values obtained by RAPID when increasing the number of subjects from $2$ to $7$.
The obtained values are not significantly lower from those obtained with \ac{mmwave} radars, and in some cases even superior, e.g., the $79\%$ on $5$ subjects in \cite{vandersmissen2018indoor}, the $98\%$ with $4$ subjects in \cite{pegoraro2021multiperson} or the $89\%$ with $12$ subjects in \cite{zhao2019mid}.
This is even more valuable considering the few available training data and the short duration of the observation window used, compared to the windows used in the mentioned papers which vary between $2$ and $3$~s.

\smallsection{Continuous \ac{har} and person identification}
Finally, we show that RAPID is capable of simultaneously \textit{(i)} tracking subjects, \textit{(ii)} recognizing their activities, and \textit{(iii)} identifying who is performing each activity from their gait.
We perform several tests in which $2$ subjects, \textit{concurrently} present in the room, perform various activities sequentially, e.g., walking then sitting, etc.
In this scenario, people tracking is of key importance to collect the temporal evolution of each subject's \ac{md}, so that all the activities performed by a person can be associated to that person's identity, obtained by RAPID when he/she is walking. 

\begin{figure}[t!]
	\begin{center}   
		\centering
		\subcaptionbox{S$1$ walking-waving. \label{fig:ftest-f2}}[4.25cm]{\includegraphics[width=4.25cm]{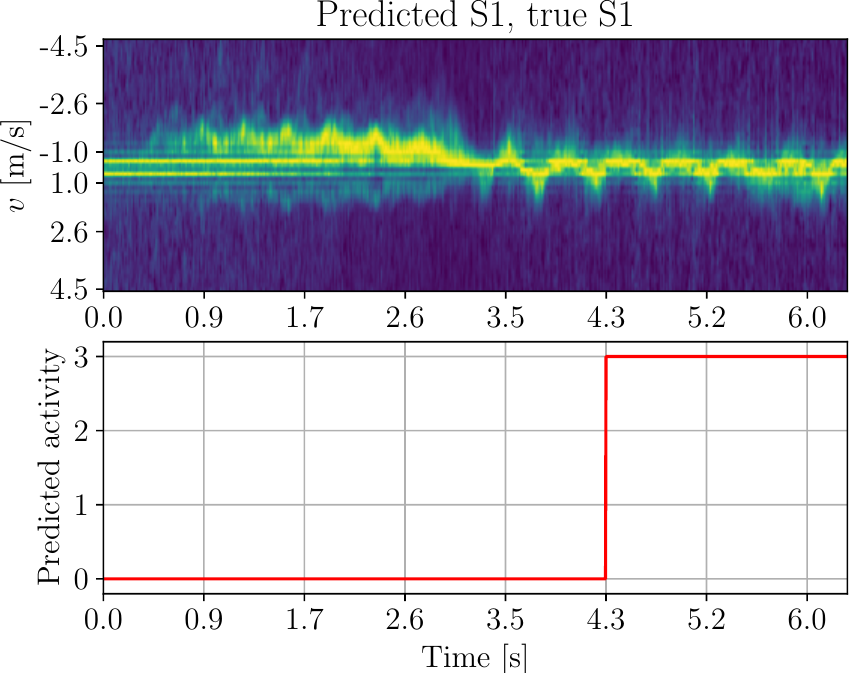}}
		\subcaptionbox{S$0$ sitting down-walking. \label{fig:ftest-f1}}[4.25cm]{\includegraphics[width=4.25cm]{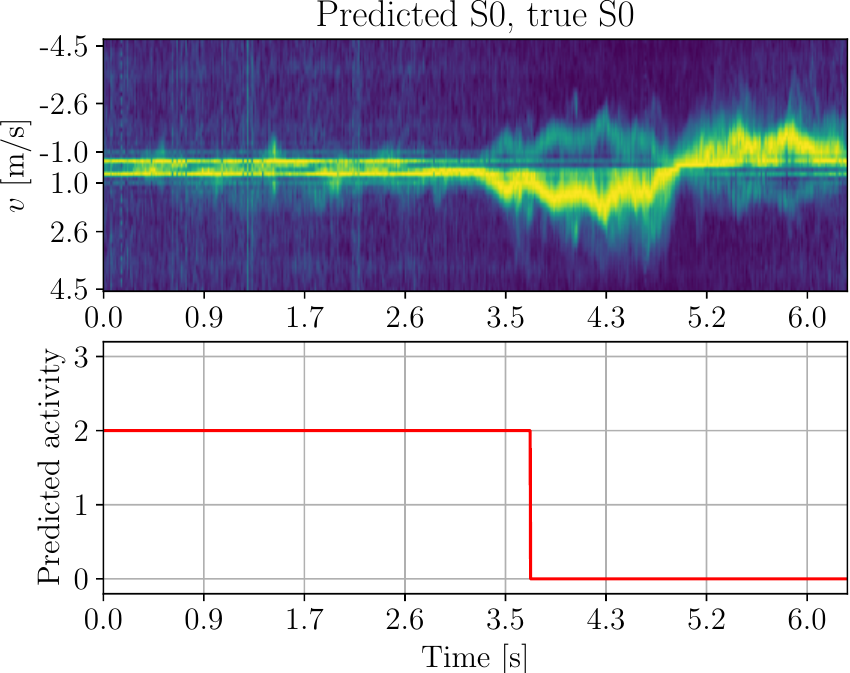}}
		\caption{\ac{md} signature and corresponding \ac{cnn} output when subject $0$ is sitting down (A$2$), then starts walking (A$0$), while subject $1$ is walking and then starts waving hands (A$3$). 
		}
    \label{fig:ftest}
	\end{center}
	\vspace{-0.6cm}
\end{figure}

In \fig{fig:ftest} we show the results obtained by RAPID with $2$ subjects, S$0$ and S$1$, behaving as follows. S$0$ enters the scene walking, then after approximately $3.5$~s S$0$ stops and starts waving hands, while S$2$ is sitting down and then starts walking after $3.5$~s. 
We report the \ac{md} signature extracted after successfully tracking the subjects, along with the predicted activity using our moving window approach.
We observe that RAPID detects the change in the activity performed by each subject; moreover, by applying the identification \ac{cnn} to the spectrogram portion where the subjects are walking, it successfully identifies them as S$0$ and S$1$ among the $7$ subjects in the training set.
\begin{table}[t!] 
	\caption{Identification accuracy vs.\ number of subjects. } \label{tab:person-id}
    \vspace{-0.25cm}
	\begin{center}
		\begin{tabular}{lcccccc}
			\toprule	
			&$2$ subj.&$3$ subj.&$4$ subj.&$5$ subj.&$6$ subj.&$7$ subj.\\
			\cmidrule(lr){2-7}
			\textbf{Acc.} $[\%]$&$97.8$&$95.9$&$94.6$&$94.1$&$92.7$&$90.0$\\
			\bottomrule
		\end{tabular} 		
	\end{center}
\end{table}

\subsection{Overhead considerations}\label{sec:overhead-cons}

The sensing operations performed by RAPID add a certain overhead to the communication process, due to: \textit{(i)} appending TRN units to the communication packets, which entails transmitting redundant bits that do not carry information; \textit{(ii)} accessing and occupying the channel for sensing, which may interfere with other communication links in the proximity of the \ac{ap}. In this section, we discuss point \textit{(i)}, while we already addressed point \textit{(ii)} in a separate work \cite{pegoraro2022sparcs}. There, we proposed a method to reconstruct \ac{md} signatures from the irregular and sparse \ac{cir} estimates obtained from communication packets, so as to minimize the need to access the channel for the sole purpose of sensing.

We can assess the overhead of RAPID by comparing the PHY layer packet size in IEEE~802.11ay to the size of TRN fields used for sensing. As shown in \fig{fig:11ay_packet}, physical layer \acp{pdu} include the \ac{stf}, the \ac{cef} and the PHY layer header,  including $\mathrm{STF}_{l}=2176$, $\mathrm{CEF}_{l}=1152$ and  $\mathrm{PHY}_{l}=1024$ samples, respectively \cite{802.11ay}. Each TRN field includes $6$ complementary Golay sequences, for a total of $\mathrm{TRN}_l = 768$~samples. Therefore, the overhead introduced by appending $\xi$ TRN fields to a packet is 
\begin{equation}
    O = \frac{\mathrm{TRN}_l \cdot \xi} {\mathrm{STF}_{l} + \mathrm{CEF}_{l} + \mathrm{PHY}_{l} + \mathrm{DATA}_l + \mathrm{TRN}_l \cdot \xi},
\end{equation} 
where $\mathrm{DATA}_l$ is the length of the data portion of the packet.
We recall that, with RAPID, it is sufficient to illuminate a person with \textit{one} \ac{bp} to apply the extraction of the \ac{md} spectrum, and that we can use one \ac{bp} per TRN field, so $\xi$ can be selected equal to the number of subjects tracked by RAPID. 
In order to reduce the inefficiency of the MAC layer and achieve Gigabit data rates, in IEEE~802.11ay large packet aggregation is permitted, allowing PHY layer \acp{pdu} to contain up to $4$~MB of data. For this, multiple MAC layer \acp{pdu} of $1.5$~kB are encapsulated into a single PHY layer packet. Compared to these large packet sizes, the TRN fields used by RAPID add a limited amount of overhead. To see this, consider that, e.g. \ac{mcs} $8$ is used, and that the data size is $20$~kB (note that is a very small fraction of the maximum allowed aggregated packet size). Even in this conservative example we get $\mathrm{DATA}_l = 126784$ samples (due to the \ac{mcs} used) \cite{802.11ay}, leading to $O=0.6\cdot \xi \%$. Moreover, RAPID does not require the TRN fields to be appended to \textit{every} PHY layer packet, but only to one every $T_c$ seconds. With $T_c=0.27$~ms as in our implementation, considering the same data size used above and the IEEE~802.11ay sample rate of $1.76$~Gsps, we get that the TRN fields need to be added to only one out of $3-4$ PHY layer \acp{pdu}, further reducing the overhead.

As a final note, we stress that RAPID performs sensing using reflections of standard-compliant packets, i.e., the transmitted packets are not designed for sensing, but we rather exploit some properties of the standard itself to enable \ac{jcr}. 
While the header and payload of the packet are transmitted with the \ac{bp} that maximizes the communications quality towards the intended receiver, the appended TRN fields can be transmitted with an arbitrary \ac{bp}. To obtain a signal reflection to be used for sensing, we use the \ac{bp} that illuminates the target.
Therefore, the sensing operations in RAPID do not interfere with communication besides the addition of TRN fields, which have a small impact on the overall throughput, as discussed above.

\section{Concluding remarks} \label{sec:conclusion}

In this paper, we have designed and implemented RAPID, the first \ac{mmwave} \ac{jcr} system performing high-resolution sensing of human \ac{md} signatures through standard-compliant IEEE~802.11ay packets. RAPID uses the in-packet TRN fields, as specified by the 802.11ay standard, to estimate the channel impulse response. This makes it possible to perform joint tracking and localization of multiple people freely moving in an indoor environment. In addition, their \ac{md} signatures are extracted by analyzing the phase difference between subsequent packets, which enables advanced sensing tasks such as continuous \ac{har} and person identification, with radar-level accuracy.
RAPID successfully combines the high-resolution sensing capabilities of \ac{mmwave} radars with the scalability and ease of deployment of existing communication hardware, allowing the seamless integration of multiple \ac{ap}s. We implemented two RAPID \ac{ap}s with full-duplex capabilities on an FPGA-based \ac{sdr} platform equipped with phased antenna arrays, and we have thoroughly evaluated the system performance through an extensive measurement campaign. Our results show that $2$ combined RAPID-\ac{ap}s can track up to $5$ subjects concurrently moving in an indoor environment, achieving accuracies of up to $94\%$ and $90\%$ for \ac{har} and person identification, respectively. Moreover, in \ac{har}, RAPID performs significantly better than standard sub-$6$~GHz sensing, showing better capability of distinguishing similar activities and generalizing to new environments and unkwnown subjects.

Future research directions includes the combination of our system with sub-$6$~GHz radios, to benefit from the points of strength of both frequency domains: while \ac{mmwave} signals are ideal for localization, tracking and \ac{md} extraction, systems operating at lower frequencies can improve the recognition of movements that do not involve a large displacement in the radial direction with respect to the receiver (e.g., sitting down), thanks to their richer multipath environment.
Other research avenues include {\it (i)} extending the RAPID system to bistatic network configurations, where sending and receiving units do not share a common phase reference, {\it (ii)} devising additional data fusion strategies for multiple \acp{ap} and {\it (iii)} assessing the sensing performance limits at \acp{mmwave} as a function of number and location of the \acp{ap}, size of the indoor space to be monitored, people density and number and type of objects in the environment, which may lead to occlusions and spurious reflections.

\bibliography{references}
\bibliographystyle{ieeetr}
\vspace{-0.3cm}
\begin{IEEEbiography}[{\includegraphics[width=1in,height=1.25in,clip,keepaspectratio]{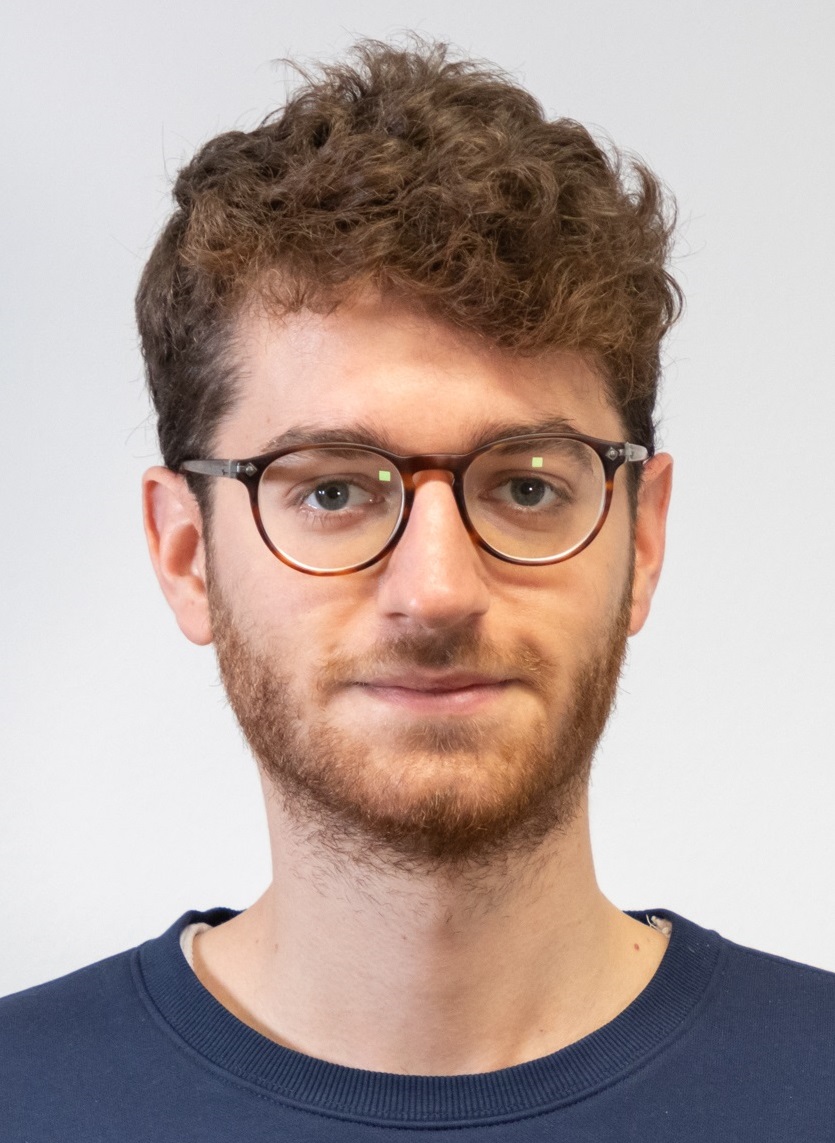}}]
{Jacopo Pegoraro} (S'20) received his Ph.D. in Information Engineering from the University of Padova, Padua, Italy, in 2023. He is currently working as a postdoctoral researcher in the Department of Information Engineering, in the same University. He was a visiting research scholar at the New York University, Tandon school of Engineering in 2022. His research interests include signal processing and machine learning for \ac{mmwave} sensing and integrated sensing and communication.
\end{IEEEbiography}
\vspace{-0.3cm}
\begin{IEEEbiography}[{\includegraphics[width=1in,height=1.25in,clip,keepaspectratio]{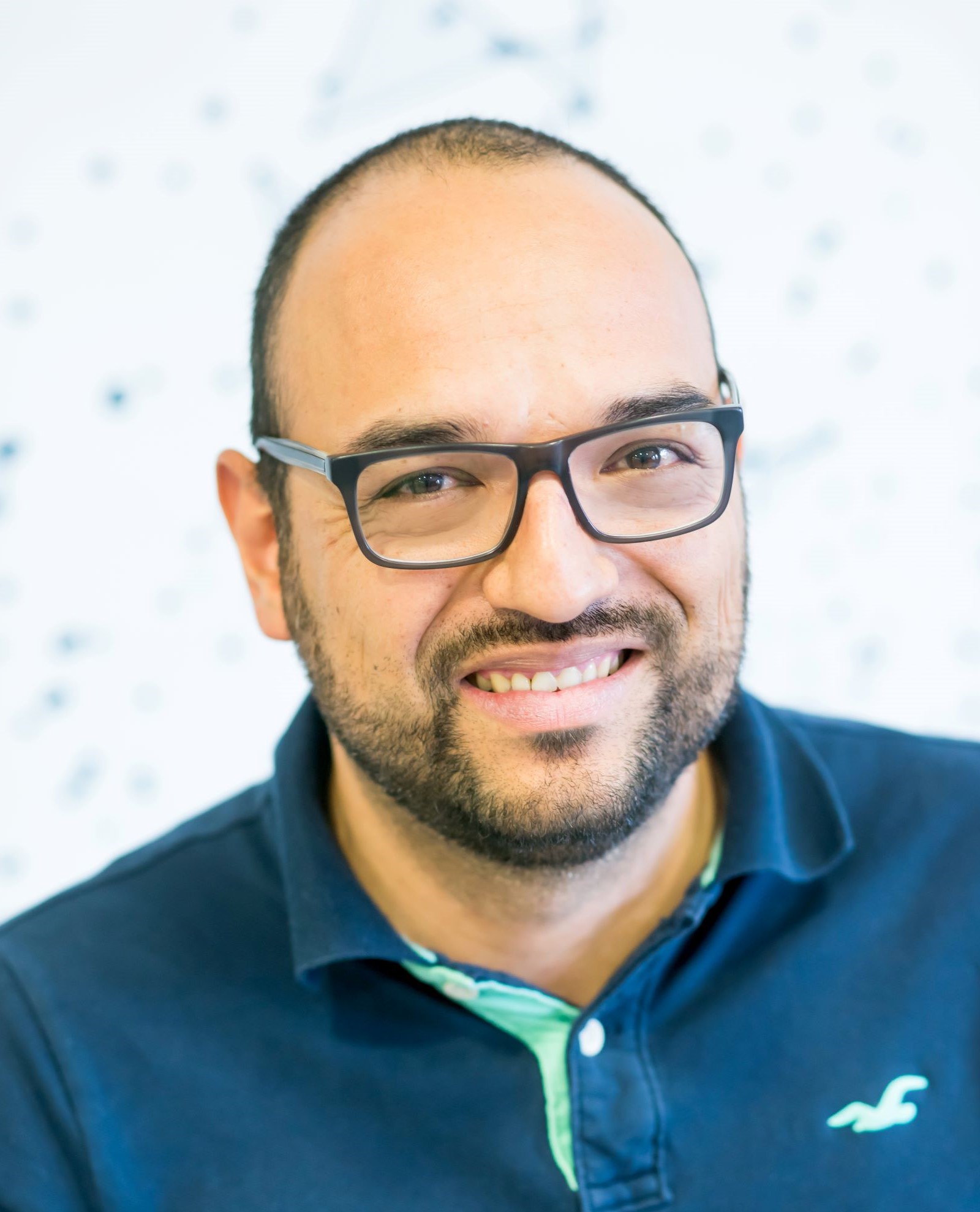}}]{Jesus O. Lacruz }
is a Research Engineer at IMDEA Networks, Spain since 2017. He received his Bachelor degree in Electrical Engineering from Universidad de Los Andes, Venezuela in 2009 and the PhD degree in Electronic Engineering from Universidad Politecnica de Valencia, Spain in 2016. His research interests lie in the design and implementation of fast signal processing algorithms for digital communication systems in FPGA devices.
\end{IEEEbiography}
\vspace{-0.3cm}
\begin{IEEEbiography}[{\includegraphics[width=1in,height=1.25in,clip,keepaspectratio]{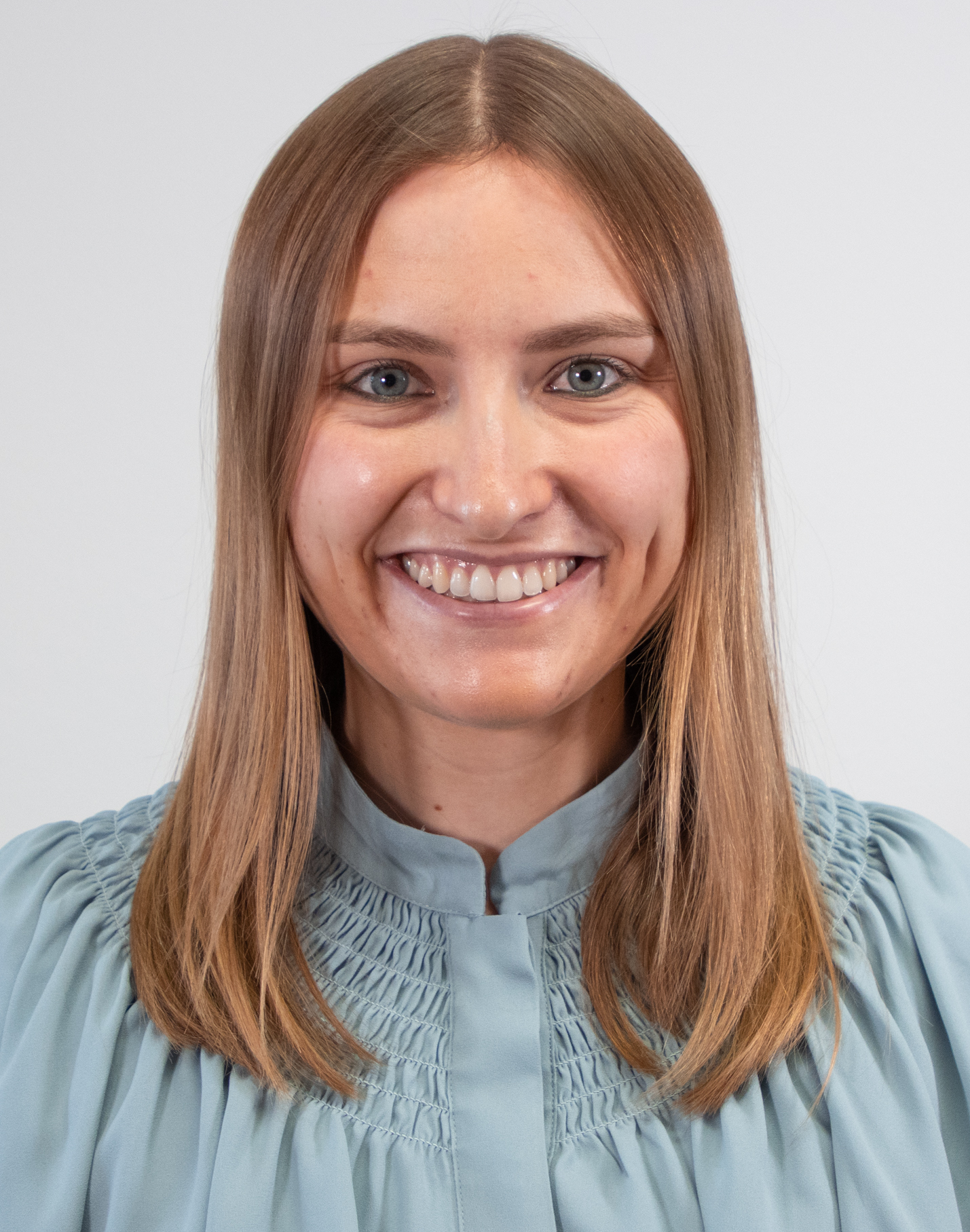}}]
{Francesca Meneghello} (S'19) received the Ph.D. degree in information engineering from the University of Padova, Italy, in 2022. She is currently an assistant professor at the Department of Information Engineering at the same university. Her current research interests include \mbox{deep-learning} architectures and signal processing with application to remote radio frequency sensing and wireless networks. 
\end{IEEEbiography}
\vspace{-0.3cm}
\begin{IEEEbiography}[{\includegraphics[width=1in,height=1.25in,clip,keepaspectratio]{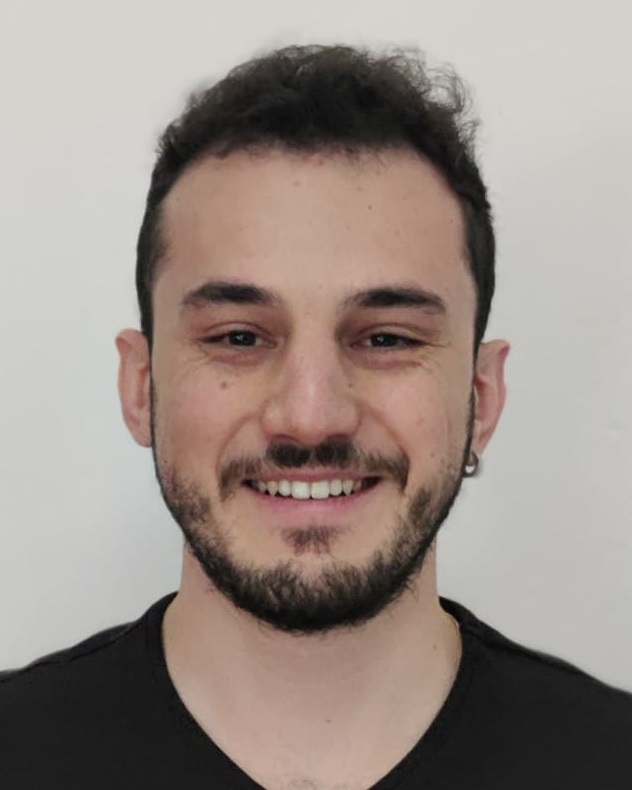}}]
{Enver Bashirov} (S'20) is currently an early-stage researcher at EU Horizon 2020 Marie Skłodowska-Curie project MINTS, pursuing his Ph.D. degree at the Department of Information Engineering, University of Padova, Italy. He received his M.Sc. degree in Applied Mathematics and Computer Science from Eastern Mediterranean University, North Cyprus. His research interests include sensing applications in \ac{mmwave}, together with machine learning and signal processing solutions.
\end{IEEEbiography}
\vspace{-0.3cm}
\begin{IEEEbiography}[{\includegraphics[width=1in,height=1.25in,clip,keepaspectratio]{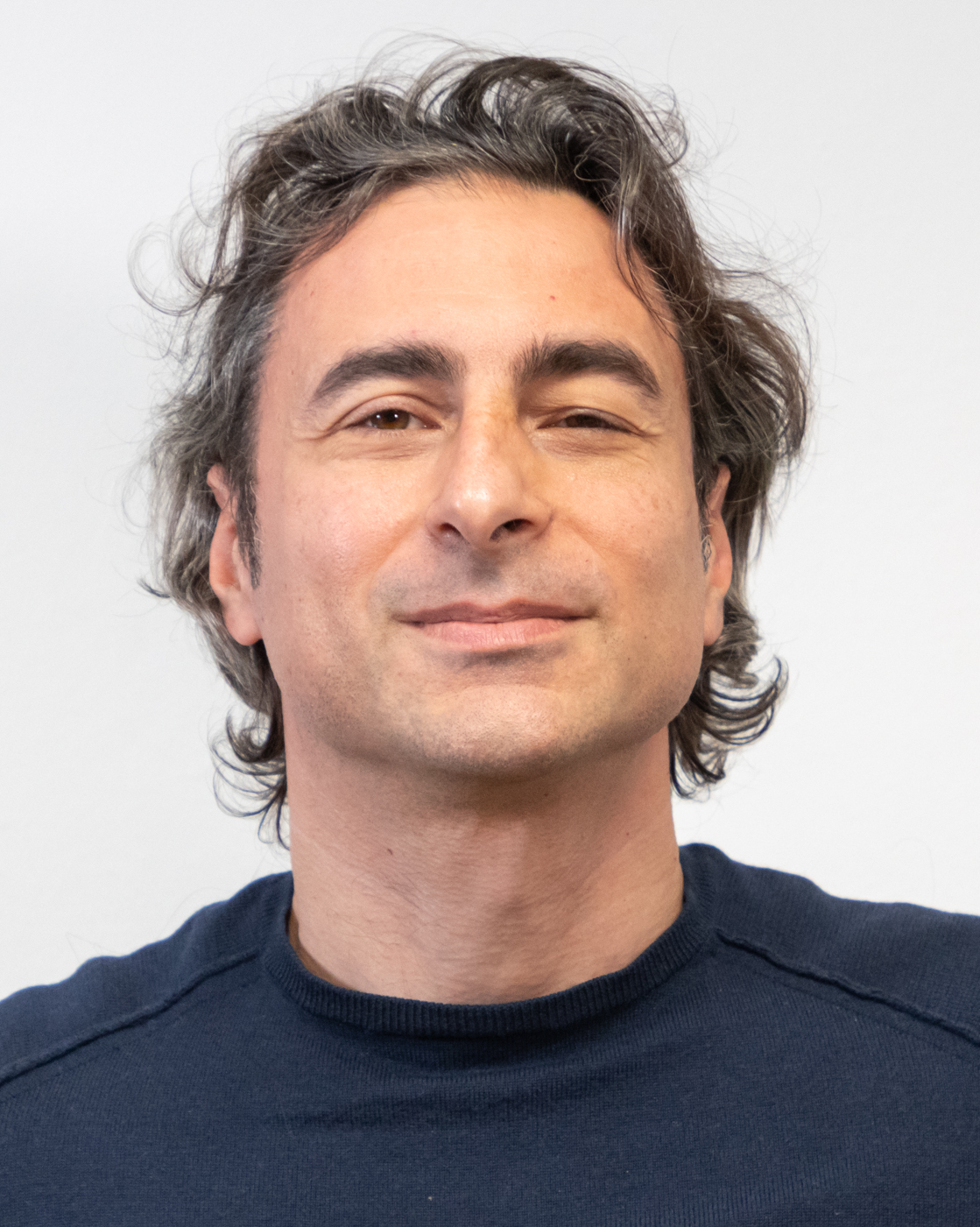}}]%
{Michele Rossi} (SM'13) is the head of the Master's Degree in ICT for internet and Multimedia (MIME) and full professor at the Department of Information Engineering of the University of Padova. Since 2017, he has been the Director of the DEI/IEEE Summer School of Information Engineering (SSIE), held yearly in Brixen, Italy. He is also the coordinator of the GREENEDGE (no. 953775) ITN project on ``green edge computing for mobile networks'’. His research interests are on wireless sensing and edge computing systems with a focus on green ICT technologies.
\end{IEEEbiography}
\vspace{-0.3cm}
\begin{IEEEbiography}[{\includegraphics[width=1in,height=1.25in,clip,keepaspectratio]{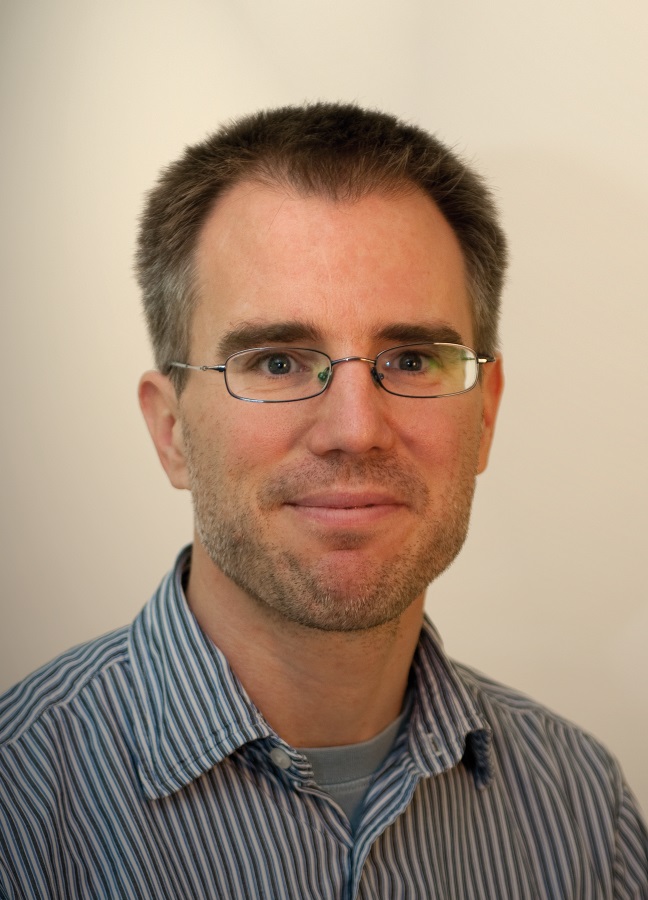}}]{Joerg Widmer}
(
F'20) 
is Research Professor and Research Director of IMDEA Networks in Madrid, Spain. 
His research focuses on wireless networks, ranging from extremely high frequency millimeter-wave communication and MAC layer design to mobile network architectures. He authored more than 150 conference and journal papers, three IETF RFCs and holds 13 patents. He was awarded an ERC consolidator grant, the Friedrich Wilhelm Bessel Research Award
, 
a Spanish Ramon y Cajal grant, as well as eight best paper awards. 
\end{IEEEbiography}

\end{document}

%% file: ACRONYMS.tex
\acrodef{prop}[\textit{MIMORPH}]{MIMO Radio Platform for Heterogeneous wireless systems}

\acrodef{abft}[A-BFT]{Association Beamforming Training}
\acrodef{ack}[ACK]{Acknowledge}
\acrodef{adc}[ADC]{Analog-to-Digital Converter}
\acrodef{aoa}[AoA]{Angle of Arrival}
\acrodef{aod}[AoD]{Angle of Departure}
\acrodef{ap}[AP]{Access Point}
\acrodef{amc}[AMC]{Advanced Mezzanine Card}
\acrodef{awv}[AWV]{Antenna Wave Vector}
\acrodef{axi}[AXI]{Advanced eXtensible Interface}
\acrodef{ber}[BER]{Bit Error Rate}
\acrodef{bft}[BFT]{Beamforming Training}
\acrodef{bp}[BP]{Beam Pattern}
\acrodef{brp}[BRP]{Beam Refinement Phase}
\acrodef{cs}[CS]{Compressed Sensing}
\acrodef{cdf}[CDF]{Cumulative Distribution Function}
\acrodef{cef}[CEF]{Channel Estimation Field}
\acrodef{cfo}[CFO]{Carrier Frequency Offset}
\acrodef{cir}[CIR]{Channel Impulse Response}
\acrodef{csi}[CSI]{Channel State Information}
\acrodef{cfr}[CFR]{Channel Frequency Response}
\acrodef{cnn}[CNN]{Convolutional Neural Network}
\acrodef{cots}[COTS]{Commercial-Off-The-Shelf}
\acrodef{dma}[DMA]{Direct Memory Access}
\acrodef{dmg}[DMG]{Directional Multi Gigabit}
\acrodef{dti}[DTI]{Data Transfer Interval}
\acrodef{ekf}[EKF]{Extended Kalman Filter}
\acrodef{elu}[ELU]{Exponential-Linear Unit}
\acrodef{fmcw}[FMCW]{Frequency-Modulated Continuous-Wave}
\acrodef{fov}[FOV]{Field-of-View}
\acrodef{gpio}[GPIO]{General Purpose Input/Output}
\acrodef{gsps}[GSPS]{Giga-Samples per Second}
\acrodef{har}[HAR]{Human Activity Recognition}
\acrodef{if}[IF]{Intermediate Frequency}
\acrodef{jcr}[JCR]{Joint Communication \& Radar}
\acrodef{los}[LOS]{Line-of-Sight}
\acrodef{lbm}[LBM]{Loop-Back Memory}
\acrodef{mae}[MAE]{Mean Absolute Error}
\acrodef{mcs}[MCS]{Modulation and Coding Scheme}
\acrodef{md}[$\mu$D]{micro-Doppler}
\acrodef{mimo}[MIMO]{Multiple Input Multiple Output}
\acrodef{mmwave}[mmWave]{Millimeter-Wave}
\acrodef{msps}[MSPS]{Mega-Samples per Second}
\acrodef{mu}[MU]{Multiple User}
\acrodef{MUSIC}[MUSIC]{MUlti SIgnal Classification}
\acrodef{nac}[NAC]{Normalized Auto Correlation}
\acrodef{nco}[NCO]{Numerical Controlled Oscillator}
\acrodef{nlos}[NLOS]{Non-Line-of-Sight}
\acrodef{pdu}[PDU]{Protocol Data Unit}
\acrodef{per}[PER]{Packet Error Rate}
\acrodef{pl}[PL]{Programmable Logic}
\acrodef{ps}[PS]{Processing System}
\acrodef{pov}[POV]{Point-of-View}
\acrodef{rf}[RF]{Radio Frequency}
\acrodef{rfsoc}[RFSoC]{Radio Frequency System on a Chip}
\acrodef{rss}[RSS]{Received Signal Strength}
\acrodef{rom}[ROM]{Read Only Memories}
\acrodef{sc}[SC]{Single Carrier}
\acrodef{sdr}[SDR]{Software Defined Radio}
\acrodef{siso}[SISO]{Single Input Single Output}
\acrodef{sls}[SLS]{Sector Level Sweep}
\acrodef{snr}[SNR]{Signal-to-Noise Ratio}
\acrodef{soc}[SoC]{System on a Chip}
\acrodef{spb}[SPB]{Signal Processing Blocks}
\acrodef{srrc}[SRRC]{Square-Root-Raised-Cosine}
\acrodef{ssr}[SSR]{Super Sample Rate}
\acrodef{sta}[STA]{Station}
\acrodef{stf}[STF]{Short Training Field}
\acrodef{su}[SU]{Single User}
\acrodef{tf}[TF]{Time-Frequency}
\acrodef{toa}[ToA]{Time of Arrival}
\acrodef{usrp}[USRP]{Universal Software Radio Peripheral}
\acrodef{vht}[VHT]{Very High Throughput}
\acrodef{wlan}[WLAN]{Wireless Local Area Network}

%% file: FIGURES/CFO.tikz
%
%
\definecolor{mycolor1}{rgb}{0.00000,0.44700,0.74100}%
\begin{tikzpicture}

\begin{axis}[%
width=0.17*4.521in,
height=0.15*3.566in,
at={(0.537in,0.413in)},
scale only axis,
xmin=0,
xmax=12,
xlabel style={font=\color{white!15!black}},
xlabel={\scriptsize Time (s)},
ymin=-171.845521726115,
ymax=171.845521726115,
ylabel style={font=\color{white!15!black}},
ylabel={\scriptsize CFO (Hz)},
yticklabel style={font=\scriptsize},
xticklabel style={font=\scriptsize},
axis background/.style={fill=white},
xmajorgrids,
ymajorgrids,
legend style={font=\scriptsize,legend cell align=left, align=left, draw=white!15!black}
]
\addplot [color=mycolor1]
  table[row sep=crcr]{%
0	-4.58923695008321\\
0.1	15.5185284404631\\
0.2	-12.5934049503057\\
0.3	-21.4718547181371\\
0.4	13.421694642098\\
0.5	-6.70981392737511\\
0.6	-17.5407244607919\\
0.7	27.7424776170757\\
0.8	-11.1460043580617\\
0.9	2.67900838771779\\
1	-22.4069224709582\\
1.1	12.287520820382\\
1.2	13.4559216408405\\
1.3	5.74649750797693\\
1.4	6.87123540421793\\
1.5	3.9985561048127\\
1.6	-10.4193961436796\\
1.7	-2.15848819123576\\
1.8	24.2304185461959\\
1.9	-14.4072947536848\\
2	-26.7460591585195\\
2.1	-11.0066319317144\\
2.2	-16.7951927660424\\
2.3	14.3968610769957\\
2.4	19.9561319429762\\
2.5	-12.8916582014103\\
2.6	-11.1783578156804\\
2.7	-29.0549590089172\\
2.8	-18.7801672776301\\
2.9	-8.44609308337209\\
3	-0.315024979157956\\
3.1	-13.0973369539679\\
3.2	-5.3517673739272\\
3.3	-6.78534318015504\\
3.4	-21.1558152863784\\
3.5	-12.2512904153482\\
3.6	0.518579614557007\\
3.7	-12.2906030675976\\
3.8	26.9276853876087\\
3.9	-6.64945157469272\\
4	-19.9055121943448\\
4.1	-9.10989897930089\\
4.2	-15.2379922661585\\
4.3	-9.77954249914227\\
4.4	-16.5686299842928\\
4.5	20.6652952179771\\
4.6	-15.0839798332588\\
4.7	15.9707629977164\\
4.8	-10.2748729853784\\
4.9	14.2926183204579\\
5	-0.76431919525888\\
5.1	2.03099183585148\\
5.2	21.173044162544\\
5.3	8.38562571570331\\
5.4	9.07900984405057\\
5.5	0.0969429020112996\\
5.6	14.3152803432527\\
5.7	11.5873135099376\\
5.8	13.388380163033\\
5.9	0.190993132938813\\
6	9.73919003741065\\
6.1	14.1843075247339\\
6.2	3.48939874415236\\
6.3	-9.48807310023182\\
6.4	-23.7711437820531\\
6.5	14.8115631823374\\
6.6	-3.307442005207\\
6.7	-17.1073870613265\\
6.8	41.7919174789865\\
6.9	13.6616148871525\\
7	15.7907736029678\\
7.1	-31.1518700872104\\
7.2	18.3397692221283\\
7.3	-0.746257761765693\\
7.4	-7.92341402755013\\
7.5	4.93272441581733\\
7.6	-15.3276819471719\\
7.7	31.8266763418729\\
7.8	-9.14770813681011\\
7.9	16.4895969413586\\
8	5.48966777488151\\
8.1	22.9933633417939\\
8.2	18.2902852865629\\
8.3	10.9324579707711\\
8.4	24.7634481805673\\
8.5	3.20991412119863\\
8.6	14.7260436835048\\
8.7	-41.4888761914502\\
8.8	19.4159135804134\\
8.9	-2.17008872818552\\
9	6.66623905777843\\
9.1	-24.8034377914338\\
9.2	-5.22900751481173\\
9.3	7.02425900982931\\
9.4	-39.0678624678843\\
9.5	-21.5403021490164\\
9.6	-20.4322470886576\\
9.7	21.6329796441781\\
9.8	42.9613804315286\\
9.9	10.0498482175165\\
10	-11.4964058943932\\
10.1	22.0844557727008\\
10.2	2.7034189795896\\
10.3	32.4382153385894\\
10.4	17.3038734272558\\
10.5	-17.8261600474774\\
10.6	4.02506875112659\\
10.7	-6.33595601180175\\
10.8	5.78734523022959\\
10.9	-3.95730776711404\\
11	-28.722227705174\\
11.1	9.4196687104108\\
11.2	30.5875954163454\\
11.3	-32.5375739484394\\
11.4	25.0634616544636\\
11.5	23.5500459848834\\
11.6	13.6910096872035\\
11.7	-12.7905824963649\\
11.8	14.0525441207334\\
11.9	16.8684748013732\\
12	18.1838738381279\\
};

\end{axis}

\end{tikzpicture}%